\documentclass[12pt]{article}
\usepackage{epsf}
\usepackage{graphicx}
\usepackage{eepic}
\usepackage{color}
\usepackage[colorlinks]{hyperref}
\usepackage{latexsym,epsfig}
\newcommand{\be}{\begin{equation}}
\newcommand{\ee}{\end{equation}}
\newcommand{\ba}{\begin{array}{c}}
\newcommand{\ea}{\end{array}}
\newcommand{\bqa}{\begin{eqnarray}}
\newcommand{\eqa}{\end{eqnarray}}

\begin{document}

\begin{center}
{\Large\bf A Dispersive Analysis on the $f_0(600)$ and $f_0(980)$
Resonances in $\gamma\gamma\to\pi^+\pi^-, \pi^0\pi^0$ Processes }
\\[10mm]
{\sc Yu~Mao,$^1$ Xuan-Gong Wang,$^1$ Ou Zhang,$^1$ H.~Q.~Zheng,$^1$
Z.~Y.~Zhou$^2$}
\\[2mm]

\noindent {\it 1: Department of Physics, Peking University, Beijing
100871, China}
\\
\noindent {\it 2:Department of Physics, Southeast University,
Nanjing 211189, China }
\\[5mm]
\today
\end{center}

\vskip 1cm
\begin{abstract}
We estimate the di-photon coupling of $f_0(600)$, $f_0(980)$ and
$f_2(1270)$ resonances in a coupled channel dispersive approach. The
$f_0(600)$ di-photon coupling is also reinvestigated using a single
channel $T$ matrix for $\pi\pi$ scattering with better analyticity
property, and it is found to be significantly smaller than that of a
$\bar qq$ state. Especially we also estimate the di-photon coupling
of the third sheet pole located near $\bar KK$ threshold, denoted as
$f_0^{III}(980)$.
 It is argued that this third sheet pole may be
originated from a coupled channel Breit-Wigner description of the
$f_0(980)$ resonance.
\end{abstract}


\section{Introduction}

The study on the $\gamma\gamma\to \pi\pi$ process has received
renewed interests recently on both experimental~\cite{belleExp} and
theoretical side.~\cite{Penn08} One major motivation to drive the
studies on the process is to extract the two photon coupling of the
resonant states appearing in the reaction, which is helpful in
exploring the underlying structure of these  (often weird)  states,
as emphasized by Pennington.~\cite{Penn072} An incomplete list of
theoretical studies along this line may be found in
Refs.~\cite{Morgan90} -- \cite{Mennessier08}.

In this paper we will also devote to the study on the
$\gamma\gamma\to \pi^+\pi^-, \pi^0\pi^0$ processes. We proceed with
previous calculations, but now take into account the coupled channel
effects in the most interesting $IJ=00$ channel.
Section~\ref{chap1:sample} reviews the coupled channel formalism
needed in this study, including: spectral function representation,
analytic continuation of coupled channel $\gamma\gamma$ scattering
amplitudes, and the dispersive integral representation of
the latter. 
  In section~\ref{sec_Born} we review the
basic technique of partial wave expansion and how to estimate the
Born term contributions. In section~\ref{sec_T} we devote to a
redetermination to the coupled channel $\pi\pi, \bar KK$ scattering
$T$ matrices  previously proposed,~\cite{AMP} with the help of new
experimental inputs at very low energies.~\cite{newke4}
Section~\ref{sec_num} is devoted to the numerical analysis on the
$\gamma\gamma\to\pi^+\pi^-,\pi^0\pi^0$ processes, both in coupled
channel and in single channel formalism. Finally in
section~\ref{sec_f} we draw conclusions based on the numerical
studies.

\section{Couple channel formalism for the $\gamma\gamma\longrightarrow\pi\pi, \bar KK$ processes}
\label{chap1:sample}
\subsection{Spectral function representations}
In the coupled channel case the relation between (the 2 by 2) $S$
matrix and  $T$ matrix is
 \begin{equation}
   S=1+2i\rho^{1/2}T\rho^{1/2}\ ,\ \ \
   \end{equation}
   where $\rho=diag(\rho_1,\rho_2)$ and
   $\rho_{n}=\sqrt{1-4m_{n}^{2}/s}$. Here subscripts $n=1,2$ represent
   $\pi\pi,\bar KK$ channels, respectively.

Let $F_1, F_2$ be the $\gamma\gamma\longrightarrow \pi\pi, \bar KK$
amplitudes, respectively, $F\equiv (F_1,F_2)^T$. Define
   $
    F^{\pm}(s)=\lim_{\epsilon\rightarrow0^+}F(s\pm i\epsilon)\ ,$
 then $F$
obeys the following spectral representation in the physical region:
  \begin{eqnarray}\label{Im1}
  \mathrm{Im}F_{1}&=&F_{1}\rho_{1}T_{11}^{*}+F_{2}\rho_{2}T_{12}^{*}\ ,\nonumber\\
  \mathrm{Im}F_{2}&=&F_{1}\rho_{1}T_{21}^{*}+F_{2}\rho_{2}T_{22}^{*}
  \end{eqnarray}
  above $\bar KK$ threshold and
  \begin{eqnarray}\label{Im2}
  \mathrm{Im}F_{1}&=&F_{1}\rho_{1}T_{11}^{*}\ ,\nonumber\\
  \mathrm{Im}F_{2}&=&F_{1}\rho_{1}T_{21}^{*}
  \end{eqnarray}
when $s$ lies between first and second threshold. From
Eqs.~(\ref{Im1}), (\ref{Im2}), along positive real axis, one has a
shorthand expression,
 \be
  F^+ =\left(
   \begin{array}{cc}
   F_{1}^{+}\\
   F_{2}^{+}
   \end{array}
   \right)=\tilde S F^-\equiv\left(
   \begin{array}{cc}
   1+2i\rho_{1}T_{11}\theta_{1}&2i\rho_{2}T_{12}\theta_{2}\\
   2i\rho_{1}T_{21}\theta_{1}&1+2i\rho_{2}T_{22}\theta_{2}
   \end{array}
   \right)
   \left(
   \begin{array}{cc}
   F_{1}^{-}\\
   F_{2}^{-}
   \end{array}
   \right)
  \ee
where $\theta_1=\theta(s-4m_\pi^2)$ and $\theta_2=\theta(s-4m_K^2)$
are step functions.

From above spectral function representations, the analytic
continuation of $F$ to different sheets can be obtained:
 \bqa\label{B2}
 { F}^{II} &=& { B}_{II}{F} \equiv
\left(\matrix{{1 \over 1+2i\rho_1T_{11}^I},& 0
 \cr {-2i\rho_1T_{12}^I \over
1+2i\rho_1T_{11}^I} ,& 1 \cr }\right){F}\ ,\nonumber\\
 {F^{III}}&=& {B_{III}F  }\equiv  \left(\matrix{{1+2i\rho_2
T_{22}^I\over \mathrm{det}\, S}, &{-2i\rho_2T_{21}^I \over
\mathrm{det}\, S} \cr {-2i \rho_1 T_{12}^I\over \mathrm{det}\, S},
 & {1+2i\rho_1 T_{11}^I \over \mathrm{det}\, S}
 \cr }\right){F} \ ,\nonumber\\
{F}^{  IV}&=& {  B}_{IV}{F} \equiv  \left(\matrix{1, &
{-2i\rho_2T_{21}^I \over 1+2i\rho_2T_{22}^I}  \cr 0, & {1 \over
1+2i\rho_2T_{22}^I}
 \cr }\right){F}\ .
\eqa These expressions of analytic continuation will be used later
when extracting the residue couplings of each resonance.
\subsection{Dispersive representation for $\gamma\gamma\longrightarrow \pi\pi, \bar KK$ amplitudes} \label{sec_dis}
In this subsection we review the  method originally proposed by
Basdevant and collaborators~\cite{babelon76} on how to set up the
dispersive representation for the $\gamma\gamma\to \pi\pi$
amplitudes.

On the right hand cut, $F^{+}=\tilde{S}F^{-}$. Further let
 $\tilde{F}=F-F_{L}$, $F_{L}$ contains only left hand cut of $F$. In
 this paper we approximate $F_L$ by its Born term, $F_B$. Then on the positive real axis one has,
 \begin{eqnarray}
 &&F=\widetilde{F}+F_{B}\ ,\\
 &&F^{+}=(\tilde{F}+F_{B})^{+}=\tilde{F}^{+}+F_{B}=\tilde{S}F^{-}=\tilde{S}(\tilde{F}^{-}+F_{B})\ ,\\
 &&\tilde{F}^{+}=\tilde{S}\tilde{F}^{-}+(\tilde{S}-1)F_{B}\
 .\label{Ftilde}
 \end{eqnarray}
Next we will search for a $2\times
 2$ invertible matrix function $D(s)$, which is analytic on the cut $s$ plane and only contains right hand
 cut ($r.h.c.$), and on the $r.h.c.$ it obeys the same unitarity
 equation as $F(s)$. That is
 \begin{eqnarray}
 D^{+}(s)=\tilde{S}(s)D^{-}(s)\ ,\ \ \  \mathrm{or}\,\,\,\,\,
 \tilde{S}=D^{+}(D^{-})^{-1}\ .
 \end{eqnarray}
One can then deduce from the above equation and Eq.~(\ref{Ftilde})
that
 \begin{eqnarray}
  \mathrm{Im}D^{-1}\tilde{F}=-\mathrm{Im}D^{-1}F_{B}\ ,
 \end{eqnarray}
 from which one obtains a dispersive representation  for $\tilde
 F^+$
 \begin{eqnarray}\label{Ftilde'}
\tilde{F}^{+}=-\frac{D^{+}}{\pi}\int_{s_{0}}^{\infty}\frac{\mathrm{Im}D^{-1}F_{B}}{s'-s-i\epsilon}ds'+D^{+}P\
,
 \end{eqnarray}
where $P$ is a 2
 dimensional array of subtraction polynomial, and  a proper
 subtraction is understood on the dispersion integral.
 By making use of Low's theorem~\cite{Low54} which tells that when
$s\rightarrow 0$, $F(s)\rightarrow F_{B}(s)$, we can rewrite the
above equation with one more subtraction, and
obtain:~\cite{Donoghue}
\begin{equation}
F(s)=F_{B}+D(s)[Ps-\frac{s^{2}}{\pi}\int_{4m_{\pi}^{2}}^{\infty}
\frac{\mathrm{Im}D^{-1}(s')F_{B}(s')}{s'^{2}(s'-s-i\epsilon)}ds']\ .
\end{equation}
From the above discussion we know that the
$\gamma\gamma\rightarrow\pi\pi,\bar KK$ amplitudes can be fit with
parameter(s) $P$, once $D(s)$  and $F_B(s)$ are known.

\subsection{Solution of function $D(s)$} \label{chap2:update}
The 2 by 2 matrix function $D(s)$ only contains right hand cuts, and
satisfies the similar unitarity relations as $F$:
 \begin{eqnarray}
 &&\mathrm{Im}D_{11}=D_{11}\rho_{1}T_{11}^{*}\theta_{1}+D_{21}\rho_{2}T_{12}^{*}\theta_{2}\, , \nonumber\\
 &&\mathrm{Im}D_{21}=D_{11}\rho_{1}T_{21}^{*}\theta_{1}+D_{21}\rho_{2}T_{22}^{*}\theta_{2}\ ;\label{D1}
 \end{eqnarray}
 \begin{eqnarray}
 &&\mathrm{Im}D_{21}=D_{12}\rho_{1}T_{11}^{*}\theta_{1}+D_{22}\rho_{2}T_{12}^{*}\theta_{2}\ ,\nonumber\\
 &&\mathrm{Im}D_{22}=D_{12}\rho_{1}T_{21}^{*}\theta_{1}+D_{22}\rho_{2}T_{22}^{*}\theta_{2}\
 .\label{D2}
 \end{eqnarray}
 The above two equations have the
 same  structure as the following
  \begin{eqnarray}
 &&\mathrm{Im}\chi_{1}=\chi_{1}\rho_{1}T_{11}^{*}\theta_{1}+\chi_{2}\rho_{2}T_{12}^{*}\theta_{2}\ ,\nonumber\\
 &&\mathrm{Im}\chi_{2}=\chi_{1}\rho_{1}T_{21}^{*}\theta_{1}+\chi_{2}\rho_{2}T_{22}^{*}\theta_{2}\
 .\label{chi}
 \end{eqnarray}
 Hence searching for solutions of $D(s)$ is equivalent to
searching for two independent solutions of the equation for the two
dimensional array
 $(\chi_{1},\chi_{2})^T$. The two independent solutions can be identified as $(D_{11},D_{21})^T$
 and $(D_{12},D_{22})^T$.
 The integral equation  to solve Eq.~(\ref{chi}) is
 \begin{eqnarray}
 \chi_{1}^{(N+1)}(s)&&=\chi_1(0)+
 \frac{s}{\pi}\int_{4m_{\pi}^{2}}^{\Lambda^2}\frac{Re[\chi_{1}^{N}(s')\rho_{1}(s')T_{11}^{*}(s')\theta_{1}]}{s'(s'-s-i\epsilon)}ds'\nonumber\\
&&+\frac{s}{\pi}\int_{4m_{\pi}^{2}}^{\Lambda^2}
\frac{Re[\chi_{2}^{N}(s')\rho_{2}(s')T_{12}^{*}(s')\theta_{2}]}{s'(s'-s-i\epsilon)}ds'\ ,\nonumber\\
 \chi_{2}^{(N+1)}(s)&&=\chi_2(0)+
 \frac{s}{\pi}\int_{4m_{\pi}^{2}}^{\Lambda^2}\frac{Re[\chi_{1}^{N}(s')\rho_{1}(s')T_{21}^{*}(s')\theta_{1}]}{s'(s'-s-i\epsilon)}ds'\nonumber\\
&&+\frac{s}{\pi}\int_{4m_{\pi}^{2}}^{\Lambda^2}\frac{Re[\chi_{2}^{N}(s')\rho_{2}(s')T_{22}^{*}(s')
\theta_{2}]}{s'(s'-s-i\epsilon)}ds'\ ,
 \end{eqnarray}
 where $N$ denotes the number of steps in the  iteration, the integration is truncated at
 $\Lambda^2= 2.4\mathrm{GeV}^2$. On the
 $r.h.s.$ of above equation, we take the real part of the numerator
 in the integrand which is of great help in increasing the speed of
 convergence.~\cite{donoghue2} In the numerical calculation a convergent solution
 emerges after approximately 15 steps of iteration.

In this paper we  solve the coupled channel $D(s)$ function in the
case of $I,J=0,0$, for a given $T$ matrix.
 For the two $d$ waves and the $I=2$ $s$ wave we only use
single channel approximation, and in such simplified case the
solution to the function $D(s)$ is known as the Omn\'es solution,
 \begin{equation}
 D(s)=\exp\left(\frac{s}{\pi}\int_{4m_\pi^2}^{\infty}\frac{\delta(s')ds'}{(s'-s)s'}\right)\
 .
 \end{equation}


\subsection{Analytic continuation, pole residues and $\Gamma(f\rightarrow\gamma\gamma)$}
\label{chap2}

In order to extract pole residues on different sheets a knowledge of
analytic continuation to the complex plane is needed. Firstly, the
analytic continuation of function $D(s)$ is simple. It satisfies the
following dispersive integral equation£º
\begin{equation}
D(s)=D(0)
+\frac{s}{\pi}\int_{4m_{\pi}^{2}}^\infty\frac{D(s')\rho(s')T^{*}(s')\theta}{s'(s'-s)}ds'\
.
\end{equation}
$D(s),\rho(s),T(s),\theta$ in above can all be matrix functions in
above formula.  To calculate $D$ with complex argument $z$ on the
first Riemann sheet one only needs to replace $s\rightarrow z$ in
above integral formula. For function $F$ the analytic continuation
is more involved, in the following we discuss this topic in some
details.

  \textbf{Residues of second sheet poles:}\\

 Since in the vicinity of a pole $z_{II}$, $S(s\sim
z_{II})\simeq S'(z_{II})(s-z_{II})$ and
\begin{eqnarray}
T^{II}(\pi\pi\rightarrow\pi\pi)(s\to z_{II})=\frac{T^{I}(s\to
z_{II})}{S'(z_{II})(s-z_{II})}\\
F^{II}(\gamma\gamma\rightarrow\pi\pi)(s\to z_{II})=\frac{F^{I}(s\to
z_{II})}{S'(z_{II})(s-z_{II})}
\end{eqnarray}
and through the definition of coupling constants:
\begin{eqnarray}
T(\pi\pi\rightarrow\pi\pi)(s\to
z_{R})&=&\frac{g_{\pi}^{2}}{z_{R}-s}\ ,\nonumber\\
F(\gamma\gamma\rightarrow\pi\pi)(s\to
z_{R})&=&\frac{g_{\gamma}g_{\pi}}{z_{R}-s}\
,\label{pole_residue_def2}
\end{eqnarray}
 one obtains the expressions of
$g_{\pi}$ and $g_{\gamma}$ for second sheet poles:
\begin{eqnarray}
g_{\pi}^{2}={-}\frac{T(z_{II})}{S'(z_{II})}\ ,\ \ \
g_{\gamma}g_{\pi}=-\frac{F(z_{II})}{S'(z_{II})}\ .
\end{eqnarray}

 \textbf{Residues of third sheet poles:}\\

Denote the third sheet pole as $z_{III}$, which is the zero of
$\mathrm{det}\, S$, then
\begin{eqnarray}
\mathrm{det}\, S(s\to z_{III})&=& (\mathrm{det}\,
S)'(z_{III})(z-z_{III})\ .
\end{eqnarray}
From
\begin{eqnarray}
T^{III}&=&T^{I}B^{III} =\left(\begin{array}{cc}
 T_{11} & T_{12} \\
 T_{21} & T_{22}
 \end{array}
 \right)
 \left(\begin{array}{cc}
 \frac{1+2i\rho_{2}T_{22}}{\mathrm{det}\, S} & \frac{-2i\rho_{1}T_{21}}{\mathrm{det}\,
 S}\\
 \frac{-2i\rho_{2}T_{12}}{\mathrm{det}\, S} & \frac{1+2i\rho_{1}T_{11}}{\mathrm{det}\, S}
 \end{array}\right)\ ,
\end{eqnarray}
the residues are hence obtainable, according to
Eq.~(\ref{pole_residue_def2}):
\begin{eqnarray}
g_{\pi}^{2}=\frac{S_{22}(z_{III})}{2i\rho_1\, \mathrm{det}\,
S'(z_{III})}\ ,\,\,\, g_{K}^{2}=\frac{S_{11}(z_{III})}{2i\rho_2\,
\mathrm{det}\, S'(z_{III})}\ .
\end{eqnarray}
From Eq.~(\ref{B2}) one further gets
\begin{eqnarray}
g_{\gamma}\,g_{\pi}&=&-\frac{F_{1}^{I}(z_{III})S_{22}(z_{III})}{(\mathrm{det}\,
S)'(z_{III})}+\frac{F_{2}^{I}(z_{III})2i\rho_{2}(z_{III})T_{21}^{I}(z_{III})}{(\mathrm{det}\,
S)'(z_{III})}\ ,\\
g_{\gamma}\,g_{K}&=&\frac{F_{1}^{I}(z_{III})2i\rho_{1}(z_{III})T_{12}^{I}(z_{III})}{(\mathrm{det}\,
S)'(z_{III})}-\frac{F_{2}^{I}(z_{III})S_{11}(z_{III})}{(\mathrm{det}\,
S)'(z_{III})}\ .
\end{eqnarray}
Having obtained the two photon couplings, $g_\gamma$, the width of
 $\sigma$,
$f_{0}(980)$ and $f_{2}(1270)$ to two photons can be calculated,
\begin{equation}
\Gamma_{\gamma\gamma}^{R}(pole)=\frac{\alpha^{2}\beta_{R}|g_{\gamma}|^{2}}{4(2J+1)m_{R}}.
\end{equation}
where $\alpha=1/137$ is the fine structure constant and
$\beta_{R}=(1-4m_{\pi}^{2}/m_{R}^{2})^{1/2}$.

\section{Partial Waves and the  Born term
contributions}\label{sec_Born}
\subsection{Partial wave expansions}
For the $\gamma\gamma\rightarrow\pi\pi$ process there are two
independent helicity amplitudes $M_{++}$, $M_{+-}$. The former
corresponds to photon helicity difference $\lambda=0$ the latter
corresponds to $\lambda=2$. They contribute to the differential
cross-section as:~\cite{Morgan90}
\begin{equation}
\frac{d\sigma}{d\Omega}=\frac{\rho}{128\pi^{2}s}[|M_{++}|^{2}+|M_{+-}|^{2}]\
,
\end{equation}
where $\rho=[1-4m_{\pi}^{2}/s]^{1/2}$, and they have partial wave
expansions involving only even $J(\geq \lambda)$,
\begin{eqnarray}
M_{++}(s,\theta,\phi)=e^{2}\sqrt{16\pi}\sum_{J\geq0}F_{J0}(s)Y_{J0}(\theta,\phi)\ ,\\
M_{+-}(s,\theta,\phi)=e^{2}\sqrt{16\pi}\sum_{J\geq2}F_{J2}(s)Y_{J2}(\theta,\phi)\
.
\end{eqnarray}
With this normalization the total cross-section reads,
\begin{equation}
\sigma=2\pi\alpha^{2}\frac{\rho}{s}\sum_{J\geq\lambda}|F_{J\lambda}|^{2}\
.
\end{equation}

At very low energies the major contribution to the amplitudes comes
from one $\pi$ exchange (OPE), i.e., the Born term, as guaranteed by
Low's theorem.~\cite{Low54}
\begin{equation}
M_{+\pm}(s,t)\rightarrow M_{+\pm}^{B}(s,t),\ \ \ \ \
(s\rightarrow0,\,t\rightarrow m_{\pi}^{2})\ .
\end{equation}
Denoting $s$--channel partial wave of the Born amplitude as
$B_{J\lambda}(s)$, then Low's theorem implies
\begin{equation}
\frac{F_{J\lambda}(s)}{B_{J\lambda}(s)}\rightarrow 1,\ \ \
F_{J\lambda}(s)-B_{J\lambda}(s)\rightarrow O(s)\ .
\end{equation}
The OPE Born term amplitude for
$\gamma\gamma\rightarrow\pi^{0}\pi^{0}$ vanishes and for
$\gamma\gamma\rightarrow\pi^{+}\pi^{-}$ processes read,
\begin{equation}
M_{++}^{B}=M_{--}^{B}
=2e^{2}m^{2}(\frac{1}{m^{2}-t}+\frac{1}{m^{2}-u})\ ,
\end{equation}
\begin{equation}
M_{+-}^{B}= M_{-+}^{B}
=-\frac{2e^{2}(m^{4}-tu)}{s}(\frac{1}{m^{2}-t}+\frac{1}{m^{2}-u})\ ,
\end{equation}
 from which OPE partial wave amplitudes are obtained:~\cite{Morgan90}
\begin{eqnarray}
&&B_{00}(s)=\frac{1-\rho^{2}}{2\rho}\ln(\frac{1+\rho}{1-\rho})\ ,\\
&&B_{20}(s)=\sqrt{\frac{5}{16}}\frac{(1-\rho^{2})}{\rho^{2}}[\frac{(3-\rho^{2})}{\rho}\ln(\frac{1+\rho}{1-\rho})-6]\ ,\\
&&B_{22}(s)=\sqrt{\frac{15}{32}}[\frac{(1-\rho^{2})^{2}}{\rho^{3}}\ln(\frac{1+\rho}{1-\rho})
+\frac{10}{3}-\frac{2}{\rho^{2}}]\ .
\end{eqnarray}

At higher energies, however, OPE Born term amplitudes are not
satisfactory to describe the left cut well. One may try to improve
this by adding crossed channel  vector and axial vector meson
exchange diagrams. Nevertheless it is difficult to judge to what
extent  adding vector and axial vector exchange contributions
improve the situation. Further discussions will be made in
section~\ref{couplefit}.

\section{The coupled channel scattering $T$ matrices}\label{sec_T}

It is necessary to choose an appropriate $T$ matrix in each channel.
For the I=2 $s$--wave we use the result of Ref.~\cite{pku3}.
For the I=0 $d$-wave amplitude we use the result from
Ref.~\cite{pipiDwave}. In both these two cases we use single channel
approximation.

\begin{figure}[h]%
\begin{center}%
\mbox{\epsfxsize=70mm\epsffile{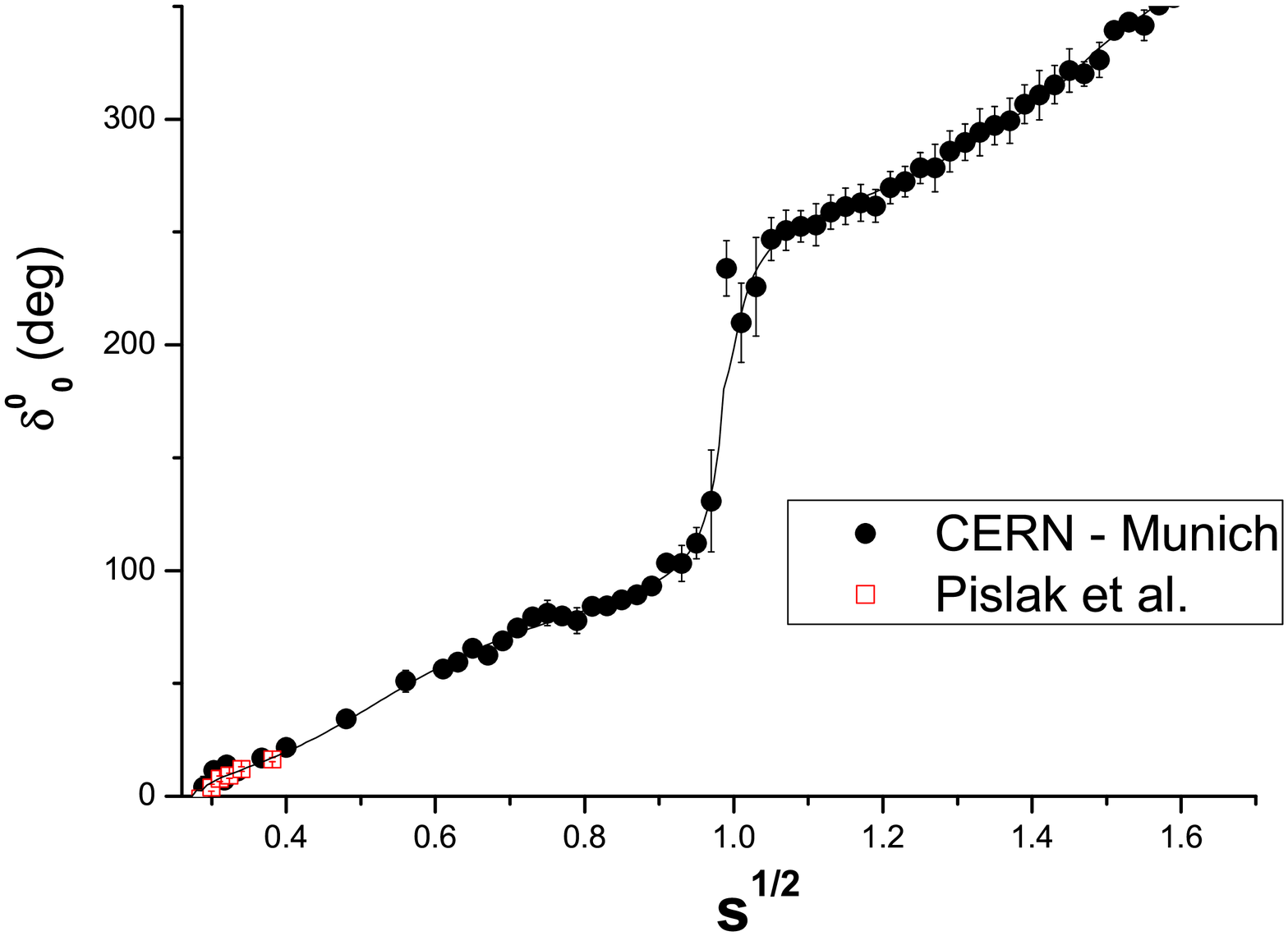}}%
\mbox{\epsfxsize=70mm\epsffile{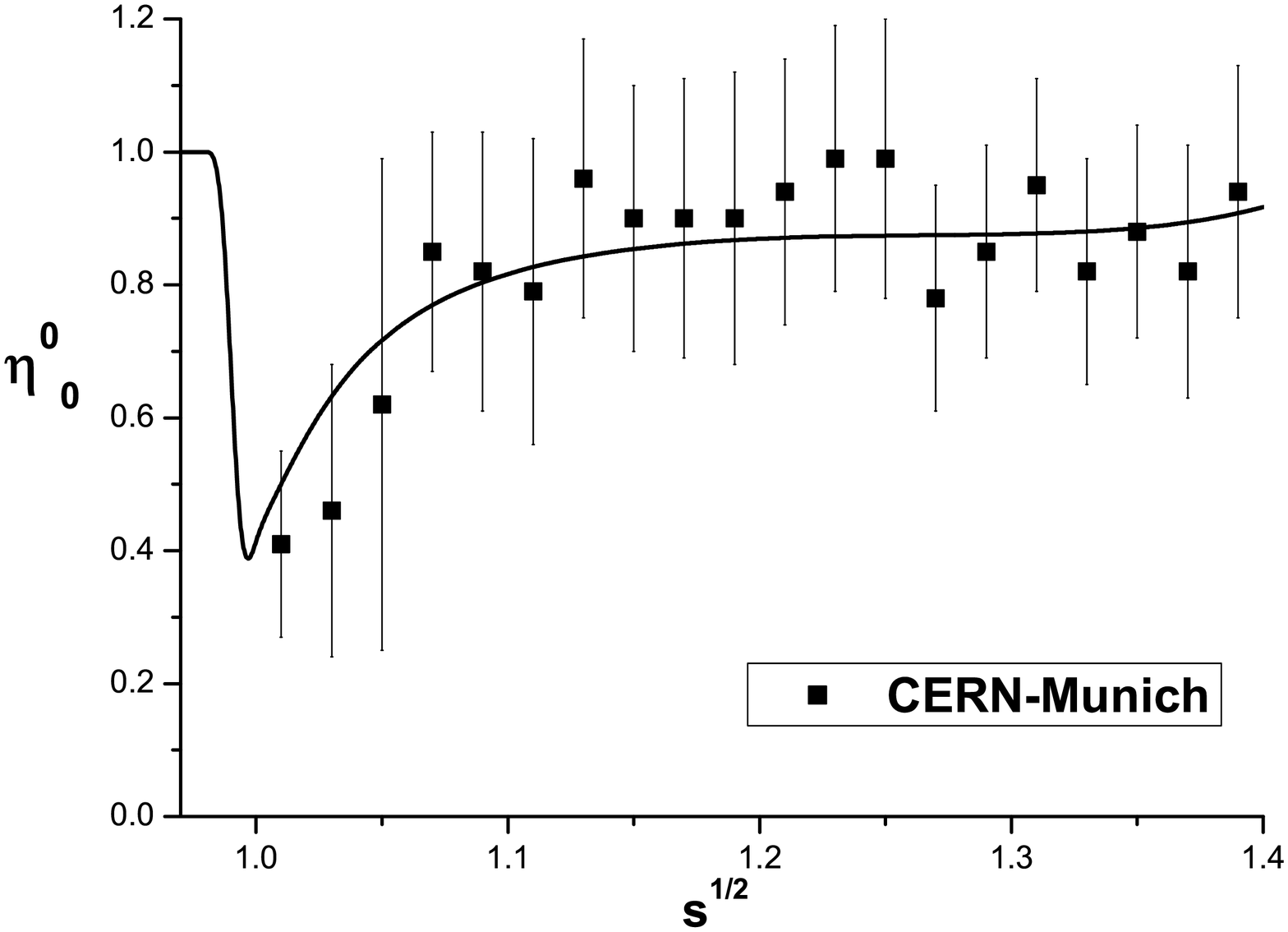}}%
\caption{\label{pipi00fit} The fit curve of $\pi\pi$
$I=0$ S-wave phase shift and inelasticity with CERN-Munich
data~\cite{Ochs} and data from Pislak et al.~\cite{newke4}.}
\end{center}%
\end{figure}

\begin{figure}[h]%
\begin{center}%
 \mbox{\epsfxsize=70mm\epsffile{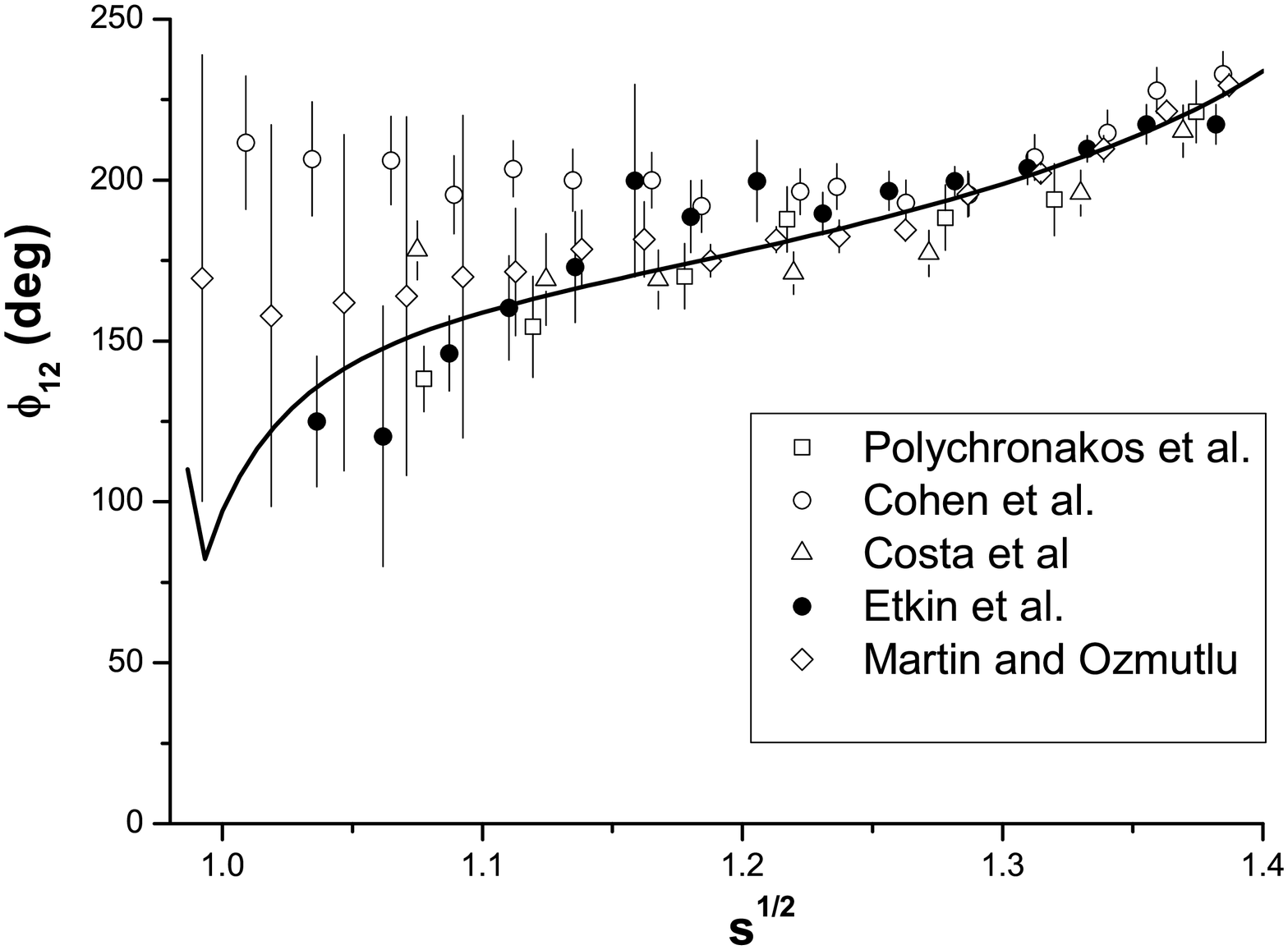}}%
\caption{\label{phase12} The phase shift
$\phi_{12}=\delta_\pi+\delta_K$ of $\pi\pi\rightarrow K\bar{K}$
$I=0$ $s$-wave scattering  with data sets from
Res.~\cite{cohen80,Etkin:1981sg,Martin:1979gm,Costa:1980ji,Polychronakos:1978ur}.
Notice that Ref.~\cite{cohen80} is not used in the fit.}
\end{center}%
\end{figure}

For the I=0 $s$-wave $\pi\pi\,\ ,\bar KK$ system, to solve the
matrix function $D(s)$ by iteration, a general coupled channel $T$
matrix is required as an input. We choose the $K$-matrix
parametrization proposed in Ref.~\cite{AMP} (the $K_3$ form) to
refit low-energy $\pi\pi$ scattering data
 with totally 22 free parameters.
The data sets we use are the phase shift and inelasticity of $I=0$
$s$-wave $\pi\pi$ scattering from CERN-Munich~\cite{Ochs} and
especially from the  $K_{e4}$
experiment~\cite{newke4},\footnote{Adding the data from
Ref.~\cite{newke4} is already good enough to constrain the $T$
matrix near threshold so the most recent NA48/2 $K_{e4}$ data is not
included.~\cite{NA48}} and $\phi_{12}=\delta_\pi+\delta_K$ of
$\pi\pi\rightarrow K\bar{K}$ $I=0$ $s$-wave scattering from several
groups.~\cite{cohen80,Etkin:1981sg,Martin:1979gm,Costa:1980ji,Polychronakos:1978ur}
  We have total 171 data points below $1.89 GeV$ in the
region of our interest. It is worthy mentioning that we can get an
acceptable fit with a $\chi^2_{d.o.f}$ of 1.14, even including the
conflicting data of $\phi_{12}$ from Cohen et al. and Etkin et al.,
but the exercise leaving aside either of them favors Etkin et al.
over Cohen et al. with a $\chi^2_{d.o.f}$ of 0.68 compared with
0.99. The final fit excluding the data from Cohen et al. are shown
in figure~\ref{pipi00fit} and figure~\ref{phase12}. The obtained
$K$--matrix parameters are listed in table~\ref{tab_K3}.
\begin{table}[hbt]
\begin{center}
\begin{tabular}{l l l}\hline\hline
 $s_{0}=0.081$            &     $c_{11}^{0}= 7.599$      &     $c_{21}^{0}=15.49$\\
 $s_{1}=0.305$            &     $c_{11}^{1}=-16.30$      &     $c_{21}^{1}=-15.60$\\
 $s_{2}=0.984$            &     $c_{11}^{2}=7.530$       &     $c_{21}^{2}=17.58$\\
 $s_{3}=3.587$            &     $c_{11}^{3}=-2.796$      &     $c_{21}^{3}=-4.942$\\
 $f_{1}^{1}=1.499$        &     $c_{11}^{4}=0.0\ \ \ $   &     $c_{21}^{4}=0.0\ \ \ $\\
 $f_{1}^{2}=1.041$        &     $c_{12}^{0}=15.49$       &     $c_{22}^{0}=-6.441$\\
 $f_{2}^{1}=0.089$        &     $c_{12}^{1}=-15.60$      &     $c_{22}^{1}=-13.53$\\
 $f_{2}^{2}=0.484$        &     $c_{12}^{2}=17.58$       &     $c_{22}^{2}=31.04$\\
 $f_{3}^{1}=-7.714$       &     $c_{12}^{3}=-4.942$      &     $c_{22}^{3}=-6.830$\\
 $f_{3}^{2}=9.216$        &     $c_{12}^{4}=0.0\ \ \ $   &     $c_{22}^{4}=0.0\ \ \ $\\
 \hline\hline
 \end{tabular}
 \caption{\label{tab_K3}$K_3$ matrix parameters obtained from the fit as described
in this section. For the definition of these
 parameters we refer to the original paper Ref.~\cite{AMP}.}
 \end{center}
 \end{table}

The fit to the $K_3$ amplitude produces several S-matrix poles
either on the second Riemann sheet or third-sheet. Those that are
relevant to the current interest are listed in
table~\ref{polelocation}.
\begin{table}[hbt]
\centering\vspace{0.1cm}
\begin{tabular}{|c|c|c|}
\hline pole& sheet--II & sheet--III
\\ \hline
$\sigma$ & $0.549-0.230i$   & $0.705-0.327i$
\\ \hline $f_0(980)$& $0.999-0.021i$  & $0.977-0.060i$ i\\
\hline
\end{tabular}
\caption{\label{polelocation}The poles's location on the
$\sqrt{s}$--plane, in units of GeV.
}
\end{table}
There we also list a third sheet pole located at $0.705-0.327i$,
which might be considered as the third sheet counterpart of
$f_0(600)$. Since it is too far away from physical region, we will
not discuss it any further. The fit with a $K_3$ parametrization
returns another pole on the third sheet, denoted as
$f_0^\mathrm{III}(980)$ herewith. It is of course a ($K_3$) model
dependent prediction. This third sheet pole is close to but below
$\bar KK$ threshold,  and may be correlated to
$f_0^\mathrm{II}(980)$,~\cite{AMP,markushin,Morgan92:pole-counting}
since the data are better described when including such a
pole.\footnote{\label{foot1}It is worthy emphasizing that the
 third-sheet pole near
$K\bar{K}$ threshold moves away and disappears, if the data of Cohen
et al. are taken into account.} We will further discuss physics
related to this pole in some detail in section~\ref{f0980BW}.

The $\sigma^{\mathrm{II}}$ pole is not very satisfactory comparing
with the determination of  Ref.~\cite{pku3}, since a coupled channel
$\pi\pi, \bar KK$ scattering amplitude fully compatible with
analyticity and crossing symmetry is still un-available.  In
section~\ref{refine} we will try to remedy this shortcoming by
making use of the single channel $T$ matrix of Ref.~\cite{pku3}. The
pole position is not very satisfactory, and the coupling strength
extracted,
 \be\label{sigmaresidue}
  g^2_{\sigma\pi\pi}=(-0.07-0.17i)\mathrm{GeV}^2 \ ,
 \ee
 differs from that given in Eq.~(\ref{sigmapipisingle}) in
 section~\ref{refine}, though the  magnitude is compatible.
On the other side it is  expected that the information one extracts
from the coupled channel is  reliable in the vicinity of $\bar KK$
threshold, especially for the $f_0(980)$ resonance.  Couplings to
$\pi\pi$ and $\bar KK$ of poles near $\bar KK$
 threshold are also obtained as listed in table~\ref{f0coupling}. It
 will be useful
  in section~\ref{f0980BW} when discussing the properties of $f_0(980)$.
 \begin{table}
\begin{center}
 \begin{tabular}  {|c|c|c|}\hline
 pole position & $g_{\pi\pi}^{2}$ & $g_{\bar
 KK}^2$  \\  \hline
$\sqrt{s_{II}}=0.999-0.021i$ &  $-0.07-0.01i$  & $-0.10+0.09i$\\
\hline
$\sqrt{s_{III}}=0.977-0.060i$ &  $-0.10+0.02i$   & $-0.02-0.09i$ \\
\hline
 \end{tabular}
 \caption{\label{f0coupling}$f_0(980)$ pole positions and their residues.}
 \end{center}
\end{table}


\section{Numerical fit to $\gamma\gamma\longrightarrow \pi^+\pi^-,\pi^0\pi^0$ process }\label{sec_num}
\subsection{The coupled channel fit}\label{couplefit}

Only I,J=0,0 channel are calculated by solving coupled channel
integral equations. Other channels are all approximated by Omn\'es
solutions. Subtraction polynomial (constant) in different channels
are listed below:
\begin{itemize}
\item $I,J,\lambda=0,0,0$: Solve couple-channel integral
equation for $D$ and fit parameters are $ P_{1}$ and $P_2$.
 \item $I,J,\lambda=2,0,0$: Use Omn\'es solution, the fit parameter
 is $P_{3}$.
 \item $I,J,\lambda=0,2,2$: Use Omn\'es function and fit
 $P_{4}$.
 \item $I,J,\lambda=2,2,2$: Since this channel's contribution is very weak, we use
 Born term approximation.
 \item $I,J,\lambda=0,2,0$: Use Omn\'es solution and the fit parameter is
 $P_{5}$.
 \item $I,J,\lambda=2,2,0$: Since this channel's contribution is very weak, we use
 Born term approximation.
 \end{itemize}
 Here five subtraction constants  are involved. Furthermore we
 introduce two additional form-factors to suppress the bad high
 energy behavior of the Born term amplitudes:
 \begin{equation}
 B_{s}\to B_{s}e^{-s/\Lambda^2_{s}}\ ,\ \ \
 B_{d}\to B_{d}e^{-s^{}/\Lambda^2_{d}}\ ,
 \end{equation}
hence totally we have 7 fit parameters.

 For $\gamma\gamma\rightarrow\pi^{+}\pi^{-}$ process there exist three sets of data,
which are from Belle,~\cite{belleExp}
 CELLO,~\cite{Behrend1992} and  Mark-II.~\cite{Boyer1990} In our fit from
 $\pi^+\pi^-$ threshold to 0.9GeV, we use the Mark-II data, from 0.9GeV to 1.4GeV we use Belle data.
For $\gamma\gamma\rightarrow\pi^{0}\pi^{0}$ there exist two data
sets from Crystal Ball Collaboration, Refs.~\cite{Marsiske1990} and
\cite{Bienlein}. Here we chose the data from
Ref.~\cite{Marsiske1990}, from $\pi\pi$ threshold up to 1.4GeV.

\begin{figure}[h]%
\begin{center}%
 \mbox{\epsfxsize=100mm\epsffile{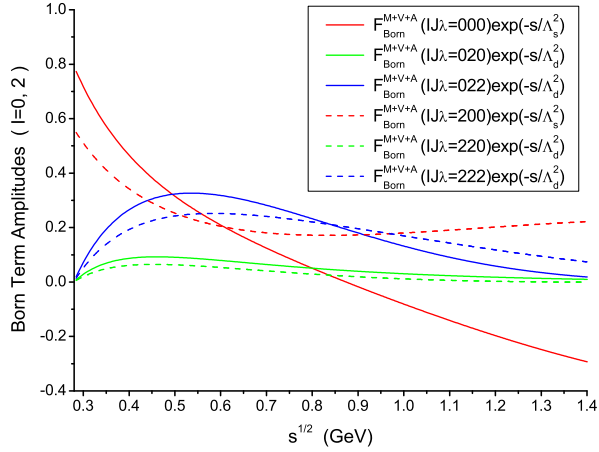}}%
\vspace{0cm} \caption{\label{Born} Born term amplitudes with
corrections from exponential  form factors taken into account. The
parameter of form-factors are determined from the fit made in
section~\ref{couplefit}.}
\end{center}%
\end{figure}

 Firstly a comparison between the contribution from OPE Born term amplitude
and the
 contribution adding vector and axial vector meson exchanges in each channel is
 made, having obtained the  two cutoff parameters which used to readjust
the Born term amplitudes: $\Lambda_s=1.77$GeV and
$\Lambda_d=0.86$GeV.
 However it is difficult to judge wether adding vector
 meson and axial vector meson exchange contributions improves
 the OPE contribution or not, comparing with the fit results shown in
 figure~\ref{Born},
 where
 contributions from one $\pi$ exchange,  vector
meson exchange,
 axial-vector meson exchange, with  corrections from the
exponential form-factors (the latter are determined from the fit),
are combined together and plotted. For the expressions of vector and
axial vector meson exchange contributions we refer to the appendix.

\begin{figure}[h]
\centering
\includegraphics[height=8cm,width=14cm]{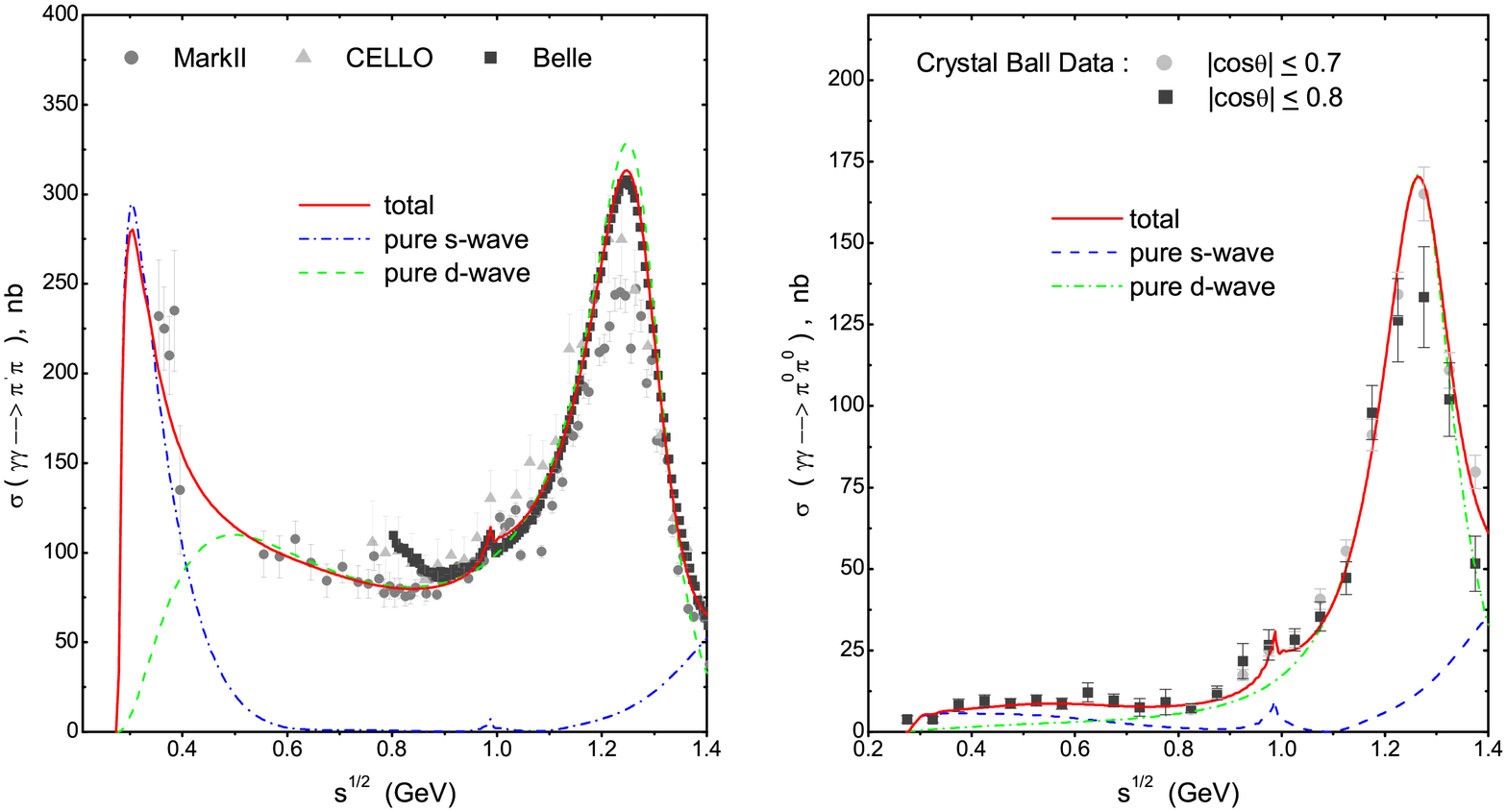}
\caption{\label{fig_couple}The coupled channel fit to the
$\gamma\gamma\to \pi^+\pi^-,\, \pi^0\pi^0$ data. The data sets
chosen are depicted in the text.}
\end{figure}

The ratio between $IJ\lambda=020$ wave and $IJ\lambda=022$ wave may
not be very stable. Though it is understood that the former should
be much smaller than the latter.~\cite{Li91} We however find a
convergent solution with $\sigma_{d0}/\sigma_{tot}\simeq 0.14$  and
$\sigma_s/\sigma_{tot}\simeq 0.07$ (at the $f_2(1270)$ pole
position).  The first ratio given here is rather close to the
`$dip$' solution of Ref.~\cite{Pennington98}. The second ratio is
significantly smaller than both the `$peak$' and the `$dip$'
solution of
 Ref.~\cite{Pennington98}. The fit results are listed in
table~\ref{table_couple} and plotted in figure~\ref{fig_couple}. The
fit result on each partial wave amplitude is found to be rather
similar in general with the solution B of Ref.~\cite{Penn08} and to
figure~1 of Ref.~\cite{achasov07}.
 The di-photon width $\Gamma(f_2\to
2\gamma)=4.36$keV here, which is larger than the value given by the
solution B of Ref.~\cite{Penn08}: $\Gamma(f_2\to
2\gamma)=3.82\pm0.30$keV.   Part of the reason for such a difference
may be due to the fact that the $s$-wave contribution given in this
paper at the $f_2(1270)$ peak is smaller than that given in
Ref.~\cite{Penn08}. Also the interference between $s$ wave and $d$
wave is found to be destructive.\footnote{Due to the limitation of
experimental detection, the integration range of $\cos\theta$ is not
from -1 to +1, hence causing a nonvanishing interference between
different partial waves.} That the result of Ref.~\cite{Penn08} on
$f_2(1270)$ di-photon width is significantly larger than the value
as quoted by PDG is explained in Ref.~\cite{Penn08} -- because the
Belle data gives a larger enhancement at $f_2(1270)$ peak than the
previous data.

\begin{table}
\begin{small}
\begin{center}
\begin{tabular}{ccc}\hline\hline
\ & Pole-positions(GeV)& $\Gamma(f_{J}\rightarrow\gamma\gamma)$(keV)\\
\hline
$ f_{0}^{II}(980)$ & $0.999-0.021i$ & $0.12$\\
$ f_{0}^{III}(980)$ & $0.977-0.060i$   & $0.35$\\
$ f_{0}(600)$ & $0.549-0.230i$    & $0.76$\\
$ f_{2}(1270)(\lambda=0)$ & $1.272-0.087i$ & $0.66$ \\
$ f_{2}(1270)(\lambda=2)$ & \  & $3.70$ \\
\hline\hline
\end{tabular}
\caption{\label{table_couple}$\chi^{2}_{d.o.f}=0.5$, using $T$
matrices as depicted in section~\ref{chap2:update}. The central
value of other 5 parameters are: $P_1= 0.25$,
 $P_2= 0.25$,
 $P_3= -0.23$,
 $P_4= 0.22$,
 $P_5= 0.55$.}
\end{center}
\end{small}
\end{table}
To summarize, we have the following observations through our fit:
\begin{enumerate}
\item
It is known that the $\lambda=2$ amplitude should dominate the
$\gamma\gamma$ width of the tensor state. Even though the ratio
between the $\lambda=0$ and the $\lambda=2$ partial wave may not be
stable, a solution with a small ratio $\sigma_{d0}/\sigma_{tot}$ is
found. The total di-photon decay width of
 $f_2(1270)$ is however found to be rather large comparing with the
 value quoted by PDG and the value given in Ref.~\cite{Penn08}.

 \item The $f_0(980)\to\gamma\gamma$ width is found to be very small. It is
 in agreement with the solution B in Ref.~\cite{Penn08}.
 See table~\ref{table_penn} for a comparison between different model
 predictions.
 \item We also give the $f^{\mathrm{III}}_0(980)$ width to two
 photons, which is not seen in previous literature.   However the third sheet pole's
 position and hence its residue
are  not quite stable, hence this number should be treated with
care.
 \item The $\sigma$ width is smaller than most values given in the
 literature.
\end{enumerate}


\subsection{A refined analysis on the sigma coupling}\label{refine}

We have noticed that the low energy physics related to the $\sigma$
pole as described by the coupled channel $T$ matrix is not very
satisfactory. Because the lack of crossing symmetry, occurrence of
spurious poles, and the distortion of the $\sigma$ pole location.
All these defects could affect a reliable extraction of the two
photon coupling of the $\sigma$ meson. We try to remedy these
defects by refitting the low energy data using a single channel $T$
matrix with all the nice properties such as analyticity, crossing
symmetry, unitarity and with a reliable $\sigma$ pole location,
since there is no coupled channel $T$ matrix with these properties
available yet. The appropriate choice is the $\pi\pi$ scattering $T$
matrix proposed in Ref.~\cite{pku3}, which gives the $\sigma$ pole
location (The Eq.~(21) of Ref.~\cite{pku3}): \be
 M_\sigma=457 MeV\ ,\,\,\, \Gamma_\sigma=551MeV\ ,
 \ee
 and the residue:
\be\label{sigmapipisingle}
g^2_{\sigma\pi\pi}=(-0.20-0.13i)\mathrm{GeV}^2 \ ,
 \ee
to be compared with Eq.~(\ref{sigmaresidue}).\footnote{The
$g^2_{\sigma\pi\pi}$ coupling differs by a factor $16\pi$ from that
of Ref.~\cite{Oller07} where the $g_{\sigma\pi\pi}\simeq 3$GeV.
Ref.~\cite{Oller07} also quoted the number extracted from
Ref.~\cite{CCL} in two ways, corresponding to
$|g_{\sigma\pi\pi}|=3.4$ and 3.9GeV, respectively.} Also the value
in Eq.~(\ref{sigmapipisingle}) is compatible with the  result of
Ref.~\cite{Mennessier08,kaminski},
$g^2_{\sigma\pi\pi}=-0.25-0.06i\mathrm{GeV}^2$.

\begin{figure}[h]
\centering
\includegraphics[height=6.5cm,width=12cm]{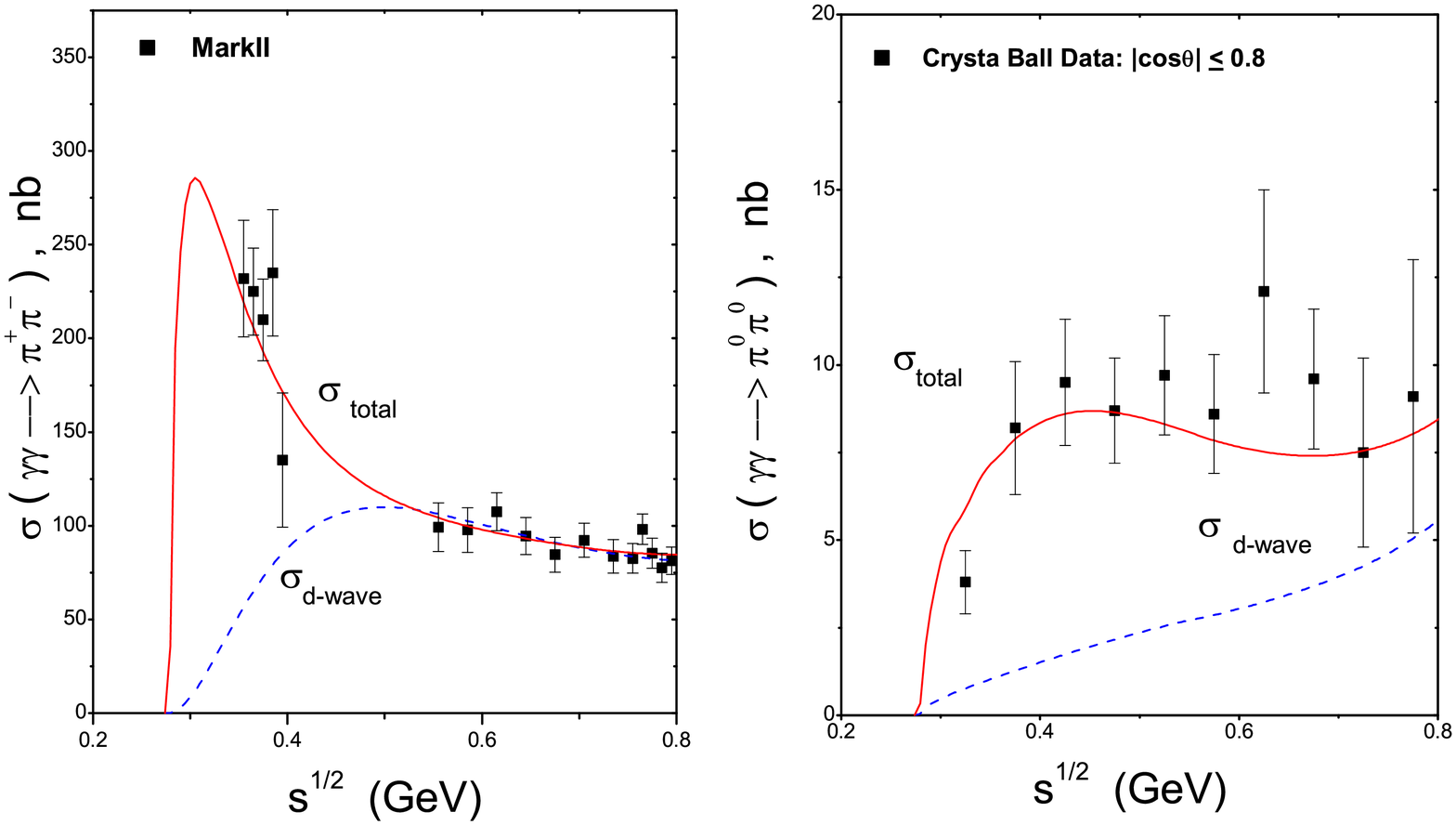}
\caption{\label{fig_single2}A fit up to 0.8GeV using single channel
$s$-wave $T$ matrices of Ref.~\cite{pku3}, with one fit parameter
$P_1$. The $\pi^+\pi^-$ and $\pi^0\pi^0$ data are from
Refs.~\cite{Boyer1990} and \cite{Marsiske1990}, respectively. Dashed
curve represents $d$-wave background, solid curve represents the
total contributions, including the I=0 $s$--wave to be fitted.
}
\end{figure}

We refit the $\gamma\gamma\to \pi^+\pi^-,\pi^0\pi^0$ data below
800MeV using that $T^{I=0}_{J=0}$ and  $T^{I=2}_{J=0}$ matrices and
keeping $d$-waves fixed by the coupled channel fit described in
section~\ref{sec_num}. That means contributions from all other
partial waves are treated as background. In this situation we   fit
with only  single parameter $P_1$, while keeping all other
parameters fixed by the coupled channel fit. The result is given in
table~\ref{single_1}. We plot the fit curve in
figure~\ref{fig_single2}.

\begin{small}
\begin{table}
\begin{center}
\begin{tabular}{c|c|c|c}\hline
$\chi^{2}_{d.o.f.}$ & $P_1$ & Pole-position (GeV) &$\Gamma(\sigma\rightarrow\gamma\gamma)$ (keV)\\
\hline
 0.8 & 0.49 & $0.457-0.276i$ & 2.08 \\
\hline
\end{tabular}
 \caption{\label{single_1}A single channel fit up to 0.8GeV data using $T$ matrix of
 Ref.~\cite{pku3}, with only one parameter $P_1$. }
 \end{center}
\end{table}
 \end{small}

\subsection{A possible Breit--Wigner description to the $f_0(980)$
resonance}\label{f0980BW}

In this section we discuss the possibility that the two narrow width
poles in the vicinity of $\bar KK$ threshold found in
section~\ref{sec_T} may be originated from a single Breit--Wigner
parametrization.

For the $\pi\pi$, $\bar KK$ coupled channel system, one can make use
of conformal mapping technique~\cite{kato} to map the four sheets
$s$ plane into the one sheet $\omega$ plane,
\begin{eqnarray}
\omega&=&\frac{1}{\Delta}(p_1+p_2)\ ,
\end{eqnarray}
where $p_i=\frac{1}{2}\sqrt{s-4m_i^2}$ denotes
 the $i$-th channel momentum, $\Delta=\sqrt{m_2^2-m_1^2}$. The $\omega $-plane is depicted in
 figure~\ref{conformal}:
\begin{figure}[h]
\centering
\includegraphics[scale=0.4]{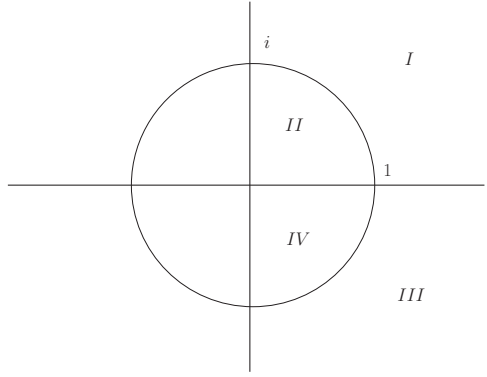}
\caption{\label{conformal}Conformal mapping  and  $\omega$--plane. }
\end{figure}
In figure~\ref{conformal} $\omega=i,1$ corresponds to $s=4m_\pi^2$,
$4m_K^2$, respectively. $S$ matrix elements can be written as,
\begin{equation}
S_{11}(\omega)=\frac{d(-\frac{1}{\omega})}{d(\omega)},\ \ \ \
S_{22}(\omega)=\frac{d(\frac{1}{\omega})}{d(\omega)},\ \ \ \ \ \det
S(\omega)=\frac{d(-\omega)}{d(\omega)}\ .
\end{equation}
Function $d(\omega)$ contains no kinematical cut and
$d(-\omega^*)=d^*(\omega)$, according to real analyticity. An  $S$
matrix pole corresponds to a zero of $d(\omega)$. We denote second
sheet pole and third sheet pole by $\omega_{2}$ and $\omega_{3}$,
respectively and
\begin{eqnarray}
\omega_{2}&=&r_2\exp{(i\phi_2)}\ ,\ \ \ \ r_2<1\ ,\,\,\,\phi_2>0\nonumber\\
\omega_{3}&=&r_3\exp{(i\phi_3)}\ ,\ \ \ \ r_3>1\ ,\,\,\,\phi_3<0\ .
\end{eqnarray}
The relations between $r_i$, $\phi_i$ and $s$-plane parameters
$s_2=(M_2-\frac{i}{2}\Gamma_2)^2$, $s_3=(M_3-\frac{i}{2}\Gamma_3)^2$
are
\begin{eqnarray}\label{pole-omega}
M_i\Gamma_i&=&(\frac{1}{r_i^2}-r_i^2)\Delta^2\sin 2\phi_i\ ,\nonumber\\
M_i^2-\Gamma_i^2/4&=&(\frac{1}{r_i^2}+r_i^2)\Delta^2\cos
2\phi_i+2(m_2^2+m_1^2)\ .
\end{eqnarray}
{In such a two-pole case,
\begin{equation}
d(\omega)=d^{bg}(\omega)D(\omega)\ ,
\end{equation}
where
\begin{equation}
D(\omega)=\omega^{-2}(\omega-\omega_2)(\omega+\omega_2^{*})
(\omega-\omega_3)(\omega+\omega_3^{*})
\end{equation}
and then
\begin{equation}\label{S}
S_{11}=\frac{D(-\omega^{-1})}{D(\omega)}\frac{d^{bg}(-\omega^{-1})}{d^{bg}(\omega)}
=\frac{D(-\omega^{-1})}{D(\omega)}S^{bg}(\omega)\ .
\end{equation}}

One can prove, {under some simplifications}, that when $r_2r_3=1$
the two pole description Eq.~(\ref{S}) is equivalent to
the following coupled channel Breit--Wigner description~\cite{Fujii}, 
\begin{eqnarray}\label{S11}
S_{11}(s)&=&\frac{s_0-s-i(-\frac{p_1}{2m_K}\gamma_{\pi}+\frac{p_2}{2m_K}\gamma_k)}
{s_0-s-i(\frac{p_1}{2m_K}\gamma_{\pi}+\frac{p_2}{2m_K}\gamma_k)}
\exp{(2i\delta_{bg})}\nonumber\\
&\simeq&\frac{s_0-s-i(-\rho_1\frac{\gamma_{\pi}}{2}+\rho_2\frac{\gamma_k}{2})}
{s_0-s-i(\rho_1\frac{\gamma_{\pi}}{2}+\rho_2\frac{\gamma_k}{2})}\exp{(2i\delta_{bg})}\
,
\end{eqnarray}
where
\begin{eqnarray}\label{BW}
s_0&=&4m_2^2+\Delta^2[(r_3-\frac{1}{r_3})^2+4\sin\phi_2\sin\phi_3]\ ,\nonumber\\
\gamma_{\pi}&=&4m_2\Delta(r_3-\frac{1}{r_3})(\sin\phi_2-\sin\phi_3)\ ,\nonumber\\
\gamma_k&=&4m_2\Delta(r_3+\frac{1}{r_3})(-\sin\phi_2-\sin\phi_3)\ .
\end{eqnarray}

The second equality in Eq.~(\ref{S11}) hold approximately, because
we limit ourselves in the energy region around $\bar KK$ threshold.
Using the pole positions provided by table~\ref{polelocation} and
Eq.~(\ref{pole-omega}) we find the pole locations on the $\omega$
plane as following:
\begin{eqnarray}
r_2&=&0.82,\ \ \ \ \sin\phi_2=0.12\ ,\nonumber\\
r_3&=&1.27,\ \ \ \ \sin\phi_3=-0.28\ ,
\end{eqnarray}
from which one finds $r_2r_3=1.04\simeq 1$. This  suggests that the
table~\ref{polelocation} does imply that $f_0^{II}(980)$ and
$f_0^{III}(980)$ may come from the same Breit--Wigner resonance. The
Breit--Wigner resonance parameters in Eq.~(\ref{S11}) can be
determined (taking $1/r_2$ as $r_3$) using Eq.~(\ref{BW}),
\begin{equation}
s_0=0.982\,\mathrm{GeV}^2,\ \ \ \gamma_{\pi}=0.146\,\mathrm{GeV}^2,\
\ \ \gamma_K=0.304\,\mathrm{GeV}^2\ .
\end{equation}
Moreover, using the information obtained from the $T$ matrix fit on
$f_0^{II}$ coupling to $\pi\pi$ (see section~\ref{sec_T}),
 \begin{equation}
(g_{\pi}^2)_{II}=(-0.071-0.011i)\mathrm{GeV}^2\ ,
\end{equation}
and comparing it with Eq.~(\ref{S11}) one can actually determine the
background phase $\delta_{bg}(s)$ at $\sqrt{s}=2m_K$ to be $\sim
91.8^\circ$,\footnote{This background phase has also been discussed
in Ref.~\cite{Bugg93}.} which is actually mainly contributed by the
$\sigma$ pole.~\cite{pku3} In this determination one make use of the
fact that $f_0$ is a narrow width resonance and hence one can assume
$\delta_{bg}(s)$ calculated at the pole position is approximately
real.

 Table~\ref{table_couple} further indicates that the $f_0^{III}$'s
di-photon  coupling is  much larger than that of $f_0^{II}$:
$|(g^2_{\gamma})_{III}/(g^2_{\gamma})_{II}|\simeq 3$. On the other
side an analysis based on coupled channel Breit-Wigner description
(Eq.~(\ref{S11})) and the two pole positions provided in
table~\ref{table_couple} gives the ratio to be $\sim 1.5$. Though
the two ratios are rather different, nevertheless they at least both
predict $|(g^2_{\gamma})_{III}|>|(g^2_{\gamma})_{II}|$.
 Hence  the two
narrow width poles in the vicinity of $\bar KK$ threshold found in
section~\ref{sec_T} may still be considered as originated from a
single Breit--Wigner parametrization, in support of an early
suggestion of Ref.~\cite{Morgan92:pole-counting}. Of course much
more efforts still need to be done to clarify such a  issue.

\section{Discussions and conclusions}\label{sec_f}

A coupled channel dispersive analysis on the $\gamma\gamma\to
\pi^+\pi^-, \pi^0\pi^0$ processes is made using a $K$ matrix
parametrization from Ref.~\cite{AMP}, but refitted with the new
$K_{e4}$ data of Ref.~\cite{newke4}. The shape of different
$\gamma\gamma\to \pi\pi$ partial wave cross-sections are similar to
those given in Refs.~\cite{Penn08,achasov07}, though we have a
smaller s wave contribution at $f_2(1270)$ peak comparing with
Ref.~\cite{Penn08}. Properties of two
 poles (one on sheet II, one on sheet III) found near $\bar KK$ threshold
 are investigated  and it
 is found that the two poles may be explained as coming from a
 single Breit-Wigner parametrization. Our prediction on the two photon width of the
 second sheet $f_0(980)$ resonance as listed in table~\ref{table_couple} is smaller comparing with
 most previous determinations found in the literature, but agrees with the solution B of Ref.~\cite{Penn08}.

 A refined analysis on the di-photon coupling of $f_0(600)$ is made,
 using the single channel $T$-matrix (the PKU) parametrization established
 in Ref.~\cite{pku3,pku1}. The PKU parametrization maintains the
 nice  property such as unitarity, analyticity and  (numerically) crossing
 symmetry.
Our result in table~\ref{single_1} gives
$\Gamma(\sigma\to2\gamma)\simeq 2.1$keV, which is compatible with
the result of Ref.~\cite{Oller07}, though in latter the
investigation only is only confined in the energy region below
0.8GeV and the $d$-wave contribution is not considered.

However, by comparing table~\ref{table_couple} and \ref{single_1},
 we find that the di-photon coupling of the $\sigma$
pole may not be very stable.  Nevertheless one may argue, based on
the analysis made in this paper, that the coupling ought to be
significantly smaller than the estimate based on a simple $(\bar
uu+\bar dd)/\sqrt{2}$ assignment. We borrow from
Ref.~\cite{Pennington07} table~\ref{table_penn} on
 radiative width of scalars in different modelings of
 their composition, adding the result from Ref.~\cite{Narison06}.
\begin{table}
\begin{center}
 \begin{tabular}  {|c||c|c|}\hline\hline
 composition & prediction& author(s) \\ \hline\hline
 $(\overline{u}u+\overline{d}d)/\sqrt{2}$&4.0&Babcock and Rosner~\cite{Rosner76}\\ \hline
 $\overline{s}s$&0.2&Barnes~\cite{Barnes85}\\ \hline
 $gg$&$0.2\sim 0.6$&Narison~\cite{Narison06}\\ \hline
 $\overline{[ns]}[ns]$&0.27&Achasov \textsl{et al}~\cite{Achasov82}\\ \hline
 $\ $&0.6&Barnes\cite{Barnes92}\\
 $\overline{K}K$&0.22&Hanhart \textsl{et al}~\cite{Hanhart07}\\ \hline\hline
 \end{tabular}
 \caption{\label{table_penn}Summary of two photon decay width of scalars calculated in different models.}
 \end{center}
\end{table}
Clearly one reads from table~\ref{table_penn} that a scalar with
di-photon decay width significantly smaller than 4keV cannot be of a
simple $\bar qq$ nature.\footnote{See however Ref.~\cite{Giacosa}.}
 It is argued in
Refs.~\cite{zheng08,guo06} that the $f_0(600)$ is the chiral partner
of the pseudo-goldstone boson, or the $\sigma$ meson responsible for
spontaneous chiral symmetry breaking. In this picture one expects
the $f_0(600)$ meson can not be described simply by a pure $\bar qq$
or a pure $\bar q^2q^2$ picture, rather it is a mixture of many
components of Fock state expansion, possibly includes a sizable glue
content as well.~\cite{Mennessier08,kaminski} It is nature then to
expect that this property is reflected by its two photon coupling as
suggested by the estimation made in this paper.

 \section{Acknowledgement}
We would like to thank Stephan Narison for helpful discussions.
Especially we are in debt to Zhi-Hui Guo for his  very valuable help
in estimating the vector meson exchange contributions. This work is
supported in part by National Nature Science Foundation of China
under Contract
Nos. 10875001, 
10721063, 
10647113 and 10705009.

\section{Appendix}
In this section we describe the method how we calculate various
vector meson and axial vector meson exchange contributions to the
Born term amplitude. We use Proca field to describe spin 1 fields.
For the $1^{--}$ vector meson, the interaction lagrangian is
\begin{equation}
\mathcal{L}_{kin}=-\frac{1}{4}<V_{\mu\nu}V^{\mu\nu}-2M_{V}^{2}V^{\mu}V_{\mu}>.
\end{equation}
Proca field is also suitable to describe the two types of axial
vector meson contributions: $\overline{A_{\mu}}(J^{PC}=1^{++})$ and
$\overline{B_{\mu}}(J^{PC}=1^{+-})$.

For $\pi\pi\gamma$ vertices,  the lowest order $\chi PT$ amplitudes
are calculable using the following interaction
lagrangian:~\cite{Gasser94}
\begin{eqnarray}
\mathcal{L}_{2}^{\chi
PT}&=&\frac{F^{2}}{4}<u^{\mu}u_{\mu}>,\nonumber\\
\mathcal{L}_{int}(V)&=&eC_{V}\varepsilon_{\mu\nu\rho\sigma}F^{\mu\nu}<V_{\rho}\{Q,u^{\sigma}\}>,
\nonumber\\
\mathcal{L}_{int}(\overline{B})&=&eC_{B}F^{\mu\nu}<\overline{B}_{\mu}\{Q,u^{\nu}\}>,
\nonumber\\
\mathcal{L}_{int}(\overline{A})&=&eC_{A}F^{\mu\nu}<\overline{A}_{\mu}[Q,u^{\nu}]>,
\end{eqnarray}
where
\begin{eqnarray}
V_{\mu}&=& \left(
   \begin{array}{ccc}
   \frac{\rho_{0}}{\sqrt{2}}+\frac{\omega}{\sqrt{2}}& \rho^{+}& K^{*+}\\
   \rho^{-}& -\frac{\rho_{0}}{\sqrt{2}}+\frac{\omega}{\sqrt{2}}& K^{*0}\\
   K^{*-}& K^{\overline{*}0}& -\phi
   \end{array}
   \right)_{\mu},\nonumber\\
\overline{A}_{\mu}&=& \left(
  \begin{array}{ccc}
  \frac{a_{1}^{0}}{\sqrt{2}}+\frac{f_{1}(1285)}{\sqrt{2}}& a_{1}^{+}& K_{1A}^{+}\\
  a_{1}^{-}& -\frac{a_{1}^{0}}{\sqrt{2}}+\frac{f_{1}(1285)}{\sqrt{2}}& K_{1A}^{0}\\
  K_{1A}^{-}& K_{1A}^{\overline{0}}& f_{1}(1420)
  \end{array}
  \right)_{\mu}, \nonumber\\
\overline{B}_{\mu}&=& \left(
  \begin{array}{ccc}
  \frac{b_{1}^{0}}{\sqrt{2}}+\frac{h_{1}(1170)}{\sqrt{2}}& b_{1}^{+}& K_{1B}^{+}\\
  b_{1}^{-}& -\frac{b_{1}^{0}}{\sqrt{2}}+\frac{h_{1}(1170)}{\sqrt{2}}& K_{1B}^{0}\\
  K_{1B}^{-}& K_{1B}^{\overline{0}}& h_{1}(1380)
  \end{array}
  \right)_{\mu} .
\end{eqnarray}

The $K_{1A}$ and  $K_{1B}$ mesons are mixed states,
\begin{eqnarray}
K_{1A}&=&\sin\theta K_{1}(1270)+\cos\theta K_{1}(1400), \nonumber\\
K_{1B}&=&-\sin\theta K_{1}(1400)+\cos\theta K_{1}(1270).
\end{eqnarray}
The mixing angle $|\theta|=37^{\circ}$ or $
58^{\circ}$\cite{Suzuki,Cheng,Guo}. Comparing with
Ref.~\cite{Gasser94}, we also introduced
$\overline{A}_{\mu}(1^{++})$ exchanges here except the $1^{--},
1^{+-}$ contributions. Coefficients $C_{V}, C_{A}, C_{B}$ are
determined through $\rho(770)\rightarrow\pi\gamma$,
$a_{1}(1260)\rightarrow\pi\gamma$ and
$b_{1}(1235)\rightarrow\pi\gamma$ processes.

 Invariant amplitudes $A, B$ are obtainable using the above interaction lagrangian:
 \newline
 \ \newline
 $\bullet  \gamma\gamma\rightarrow\pi^{0}\pi^{0}:
 \rho_{0},\omega,b_{1}^{0},h_{1}(1170)$ contributes here.
 \begin{eqnarray}
A_{N,\rho}^{t}&=&\frac{C_{V}^{2}}{9F^{2}}\frac{s-4t-4m_{\pi}^{2}}{M_{\rho}^{2}-t},
\nonumber\\
B_{N,\rho}^{t}&=&\frac{C_{V}^{2}}{18F^{2}}\frac{1}{M_{\rho}^{2}-t},\nonumber\\
A_{N,\omega}^{t}&=&\frac{C_{V}^{2}}{F^{2}}\frac{s-4t-4m_{\pi}^{2}}{M_{\omega}^{2}-t}=9A_{N,\rho}^{t}(M_{\rho}\rightarrow
M_{\omega}), \nonumber\\
B_{N,\omega}^{t}&=&\frac{C_{V}^{2}}{2F^{2}}\frac{1}{M_{\omega}^{2}-t}=9B_{N,\rho}^{t}(M_{\rho}\rightarrow M_{\omega}),\nonumber\\
A_{N,b_{1}}^{t}&=&\frac{C_{B}^{2}}{36F^{2}}\frac{s+4t-4m_{\pi}^{2}}{M_{b_{1}}^{2}-t},\nonumber\\
B_{N,b_{1}}^{t}&=&\frac{C_{B}^{2}}{72F^{2}}\frac{1}{M_{b_{1}}^{2}-t},\nonumber\\
A_{N,h_{1}}^{t}&=&\frac{C_{B}^{2}}{4F^{2}}\frac{s+4t-4m_{\pi}^{2}}{M_{b_{1}}^{2}-t}=9A_{N,b_{1}}^{t}(M_{b_{1}}\rightarrow
M_{h_{1}(1170)}),\nonumber\\
B_{N,h_{1}}^{t}&=&\frac{C_{B}^{2}}{8F^{2}}\frac{1}{M_{b_{1}}^{2}-t}=9B_{N,b_{1}}^{t}(M_{b_{1}}\rightarrow
M_{h_{1}(1170)}),
 \end{eqnarray}
 To get $u$ channel contributions we only need to replace $t$ by $u$ in above expressions,
 i.e. \ $A_{N,i}^{u}=A_{N,i}^{t}(t\rightarrow u),
 \ B_{N,i}^{u}=B_{N,i}^{t}(t\rightarrow u)$¡£\newline
 \ \newline
 $\bullet \gamma\gamma\rightarrow\pi^{+}\pi^{-}:
 \pi,\rho,b_{1},a_{1}$ contribute here.
 \begin{eqnarray}
A_{C,\pi}^{Contact}&=&\frac{4}{s},\nonumber\\
B_{C,\pi}^{Contact}&=&0,\nonumber\\
A_{C,\pi}^{t}&=&\frac{(2t+s-2m_{\pi}^{2})^{2}}{s^{2}}\frac{1}{(m_{\pi}^{2}-t)},\nonumber\\
B_{C,\pi}^{t}&=&\frac{1}{m_{\pi}^{2}-t},\nonumber\\
A_{C,\rho}^{t}&=&\frac{C_{V}^{2}}{9F^{2}}\frac{s-4t-4m_{\pi}^{2}}{M_{\rho}^{2}-t}=A_{N,\rho}^{t},\nonumber\\
B_{C,\rho}^{t}&=&\frac{C_{V}^{2}}{18F^{2}}\frac{1}{M_{\rho}^{2}-t}=B_{N,\rho}^{t},\nonumber\\
A_{C,b_{1}}^{t}&=&\frac{C_{B}^{2}}{36F^{2}}\frac{s+4t-4m_{\pi}^{2}}{M_{b_{1}}^{2}-t}=A_{N,b_{1}}^{t},\nonumber\\
B_{C,b_{1}}^{t}&=&\frac{C_{B}^{2}}{72F^{2}}\frac{1}{M_{b_{1}}^{2}-t}=B_{N,b_{1}}^{t},\nonumber\\
A_{C,a_{1}}^{t}&=&-\frac{C_{A}^{2}}{4F^{2}}\frac{s+4t-4m_{\pi}^{2}}{M_{a_{1}}^{2}-t}=-9A_{N,b_{1}}^{t}(C_{B}\rightarrow
C_{A}, M_{b_{1}}\rightarrow M_{a_{1}}),\nonumber\\
B_{C,a_{1}}^{t}&=&-\frac{C_{A}^{2}}{8F^{2}}\frac{1}{M_{a_{1}}^{2}-t}=-9B_{N,b_{1}}^{t}(C_{B}\rightarrow
C_{A}, M_{b_{1}}\rightarrow M_{a_{1}}),
 \end{eqnarray}
To get $u$ channel contribution we only need to replace $t$ by $u$
in above expressions,
 i.e.\
$A_{C,i}^{u}=A_{C,i}^{t}(t\rightarrow u),\
B_{C,i}^{u}=B_{C,i}^{u}(t\rightarrow u)$¡£\newline \
\newline
 $\bullet \gamma\gamma\rightarrow K^{+}K^{-}:
 K,K^{*},K_{1B},K_{1A}$ contribute.
 \begin{eqnarray}
A_{C,K}^{Contact}&=&\frac{4}{s},\nonumber\\
B_{C,K}^{Contact}&=&0,\nonumber\\
A_{C,K}^{t}&=&\frac{(2t+s-2m_{K}^{2})^{2}}{s^{2}}\frac{1}{(m_{K}^{2}-t)}=A_{C,\pi}^{t}(m_{\pi}\rightarrow
m_{K}),\nonumber\\
B_{C,K}^{t}&=&\frac{1}{m_{K}^{2}-t}=B_{C,\pi}^{t}(m_{\pi}\rightarrow
m_{K}),\nonumber\\
A_{C,K^{*}}^{t}&=&\frac{C_{V}^{2}}{9F_{K}^{2}}\frac{s-4t-4m_{K}^{2}}{M_{K^{*}}^{2}-t}=A_{N,\rho}^{t}(m_{\pi}\rightarrow
m_{K}, M_{\rho}\rightarrow M_{K^{*}}, F\rightarrow
F_{K}),\nonumber\\
B_{C,K^{*}}^{t}&=&\frac{C_{V}^{2}}{18F_{K}^{2}}\frac{1}{M_{K^{*}}^{2}-t}=B_{N,\rho}^{t}(m_{\pi}\rightarrow
m_{K}, M_{\rho}\rightarrow M_{K^{*}}, F\rightarrow
F_{K}),\nonumber\\
A_{C,K_{1B}}^{t}&=&\frac{C_{B}^{2}}{36F_{K}^{2}}\frac{s+4t-4m_{K}^{2}}{M_{K_{1B}}^{2}-t}=A_{N,b_{1}}^{t}(m_{\pi}\rightarrow
m_{K}, M_{b_{1}}\rightarrow M_{K_{1B}}, F\rightarrow
F_{K}),\nonumber\\
B_{C,K_{1B}}^{t}&=&\frac{C_{B}^{2}}{72F_{K}^{2}}\frac{1}{M_{K_{1B}}^{2}-t}=B_{N,b_{1}}^{t}(m_{\pi}\rightarrow
m_{K},M_{b_{1}}\rightarrow M_{K_{1B}},F\rightarrow
F_{K}),\nonumber\\
A_{C,K_{1A}}^{t}&=&-\frac{C_{A}^{2}}{4F_{K}^{2}}\frac{s+4t-4m_{K}^{2}}{M_{K_{1A}}^{2}-t}=-9A_{N,b_{1}}^{t}(C_{B}\rightarrow
C_{A}, M_{b_{1}}\rightarrow M_{K_{1A}}, m_{\pi}\rightarrow m_{K},
F\rightarrow F_{K}),\nonumber\\
B_{C,K_{1A}}^{t}&=&-\frac{C_{A}^{2}}{8F_{K}^{2}}\frac{1}{M_{K_{1A}}^{2}-t}=-9B_{N,b_{1}}^{t}(C_{B}\rightarrow
C_{A}, M_{b_{1}}\rightarrow M_{K_{1A}}, m_{\pi}\rightarrow m_{K},
F\rightarrow F_{K})\ .\nonumber\\
 \end{eqnarray}
To get $u$ channel contributions we only need to replace $t$ by $u$
in above expressions,
 i.e.\ $A_{C,i}^{u}=A_{C,i}^{t}(t\rightarrow u),
 \ B_{C,i}^{u}=B_{C,i}^{t}(t\rightarrow
 u)$¡£\newline
 \ \newline
 $\bullet \gamma\gamma\rightarrow K^{0}\overline{K}^{0}:
 K^{*},K_{1B}$ contribute.
 \begin{eqnarray}
A_{N,K^{*}}^{t}&=&\frac{4C_{V}^{2}}{9F_{K}^{2}}\frac{s-4t-4m_{K}^{2}}{M_{K^{*}}^{2}-t}=4A_{N,\rho}^{t}(m_{\pi}\rightarrow
m_{K}, M_{\rho}\rightarrow M_{K^{*}}, F\rightarrow
F_{K}),\nonumber\\
B_{N,K^{*}}^{t}&=&\frac{C_{V}^{2}}{18F_{K}^{2}}\frac{1}{M_{K^{*}}^{2}-t}=4B_{N,\rho}^{t}(m_{\pi}\rightarrow
m_{K}, M_{\rho}\rightarrow M_{K^{*}}, F\rightarrow
F_{K}),\nonumber\\
A_{N,K_{1B}}^{t}&=&\frac{C_{B}^{2}}{9F_{K}^{2}}\frac{s+4t-4m_{K}^{2}}{M_{K_{1B}}^{2}-t}=4A_{N,b_{1}}^{t}(m_{\pi\rightarrow
m_{K}}, M_{b_{1}}\rightarrow M_{K_{1B}}, F\rightarrow
F_{K}),\nonumber\\
B_{N,K_{1B}}^{t}&=&\frac{C_{B}^{2}}{18F_{K}^{2}}\frac{1}{M_{K_{1B}}^{2}-t}=4B_{N,b_{1}}^{t}(m_{\pi}\rightarrow
m_{K}, M_{b_{1}}\rightarrow M_{K_{1B}}, F\rightarrow
F_{K}),\nonumber\\
 \end{eqnarray}
To get $u$ channel contributions we only need to replace $t$ by $u$
in above expressions,
 i.e.\
$A_{C,i}^{u}=A_{C,i}^{t}(t\rightarrow u),\
B_{C,i}^{u}=B_{C,i}^{t}(t\rightarrow u)$.\newline
 In above formulas, in a term $A_{N(C),i}^{t(u)}$, the subscript $N (C)$ denote neutral (charged)
 particle exchanges.
 Superscript $i$ means resonance $R_{i}$ contribution in such channel, $t(u)$ denotes $t(u)$channel.

 One can hence get helicity amplitudes $H_{++}, H_{+-}$ for processes
  $\gamma\gamma\rightarrow\pi^{+}\pi^{-}$, $\gamma\gamma\rightarrow\pi^{0}\pi^{0}$,
 $\gamma\gamma\rightarrow K^{+}K^{-}$, $\gamma\gamma\rightarrow
 K^{0}\overline{K}^{0}$:
 \begin{eqnarray}
H_{++}&=&A+2(4m_{\pi}^{2}-s)B,\nonumber\\
H_{+-}&=&\frac{8(m_{\pi}^{4}-tu)}{s}B.
 \end{eqnarray}
For example, for the $\gamma\gamma\rightarrow\pi^{+}\pi^{-}$
process, the helicity amplitude read,
\begin{eqnarray}
H_{++}&=&A_{C,\pi}^{Contact}+A_{C,\pi}^{t+u}+A_{C,\rho}^{t+u}+A_{C,b_{1}}^{t+u}+A_{C,a_{1}}^{t+u}\nonumber\\
&&+2(4m_{\pi}^{2}-s)(B_{C,\pi}^{Contact}+B_{C,\pi}^{t+u}+B_{C,\rho}^{t+u}+B_{C_{b_{1}}}^{t+u}+B_{C,a_{1}}^{t+u}),\nonumber\\
H_{+-}&=&\frac{8(m_{\pi}^{4}-tu)}{s}(B_{C,\pi}^{Contact}+B_{C,\pi}^{t+u}+B_{C,\rho}^{t+u}
+B_{C,b_{1}}^{t+u}+B_{C,a_{1}}^{t+u})\ .
\end{eqnarray}
 The amplitudes $H_{++}, H_{+-}$ are expressed using
Condon-Shortly convention. $M_{++}, M_{+-}$ used in this paper are
defined using non Condon-Shortly convention
\begin{equation}
M_{++}=\frac{e^{2}s}{2}H_{++}\ ,\ \ \ \
M_{+-}=\frac{-e^{2}s}{2}H_{+-}\ .
\end{equation}

\begin{figure}[h]
\centering
\includegraphics[scale=0.23]{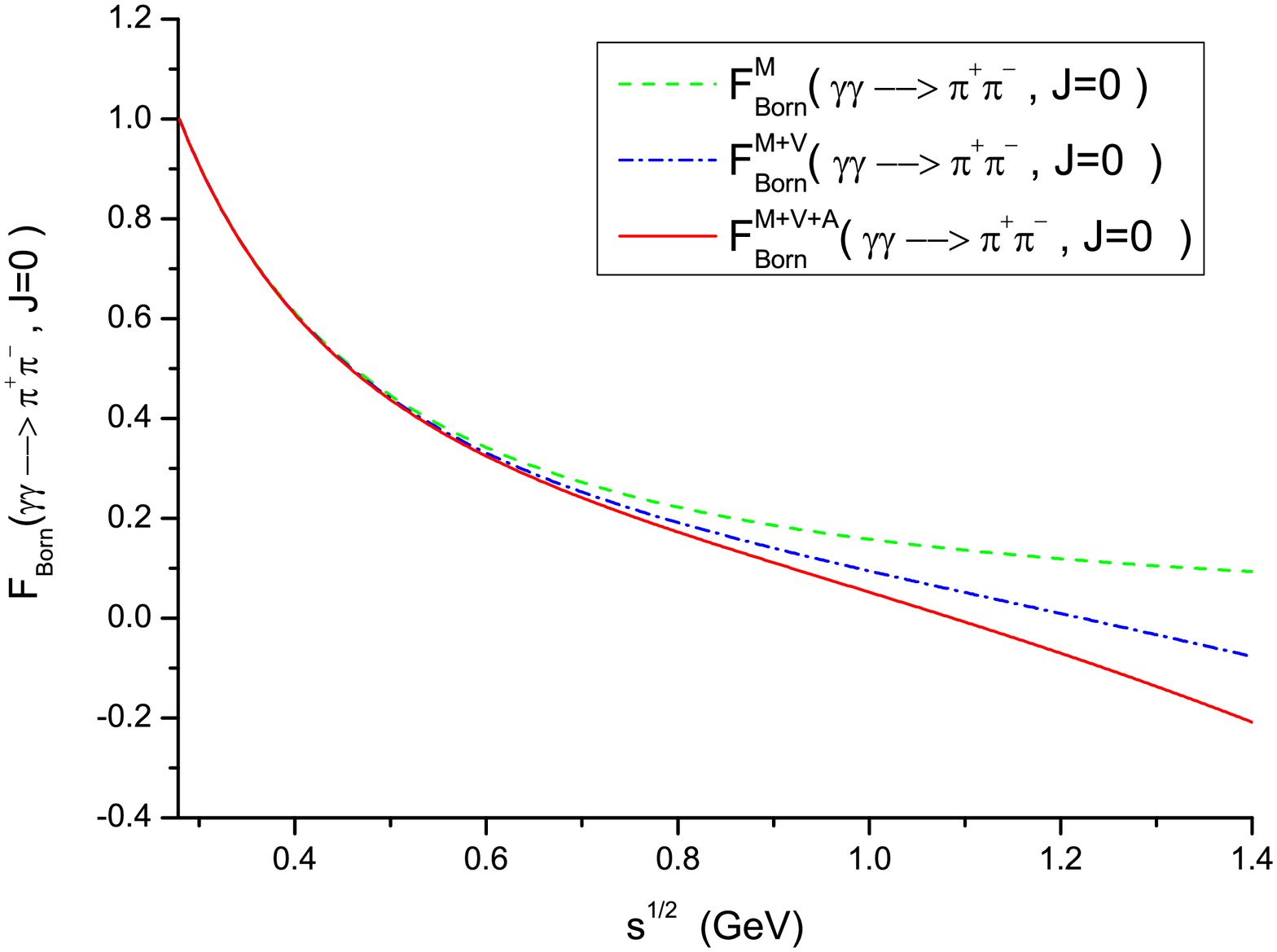}
\includegraphics[scale=0.23]{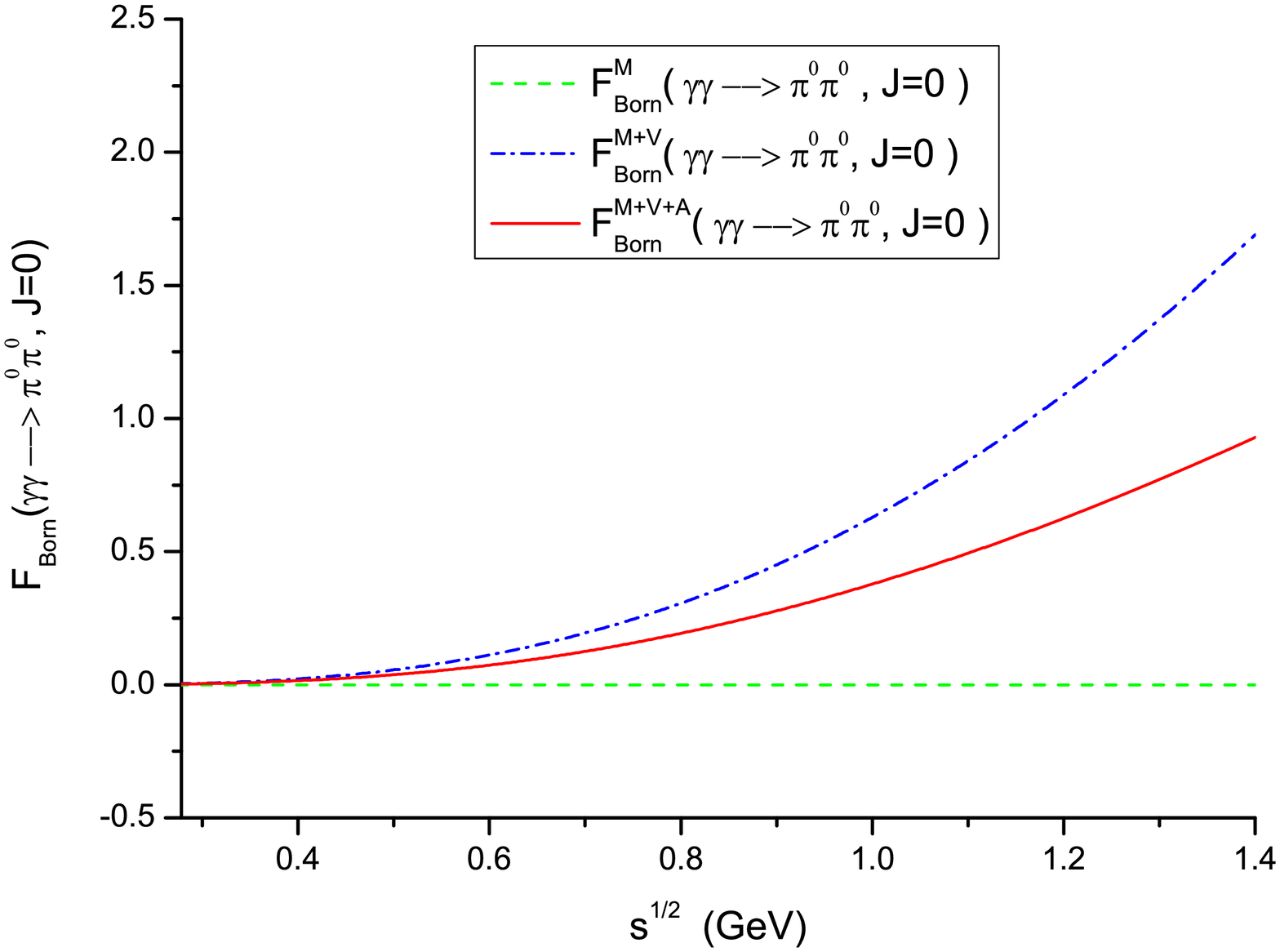}
\includegraphics[scale=0.23]{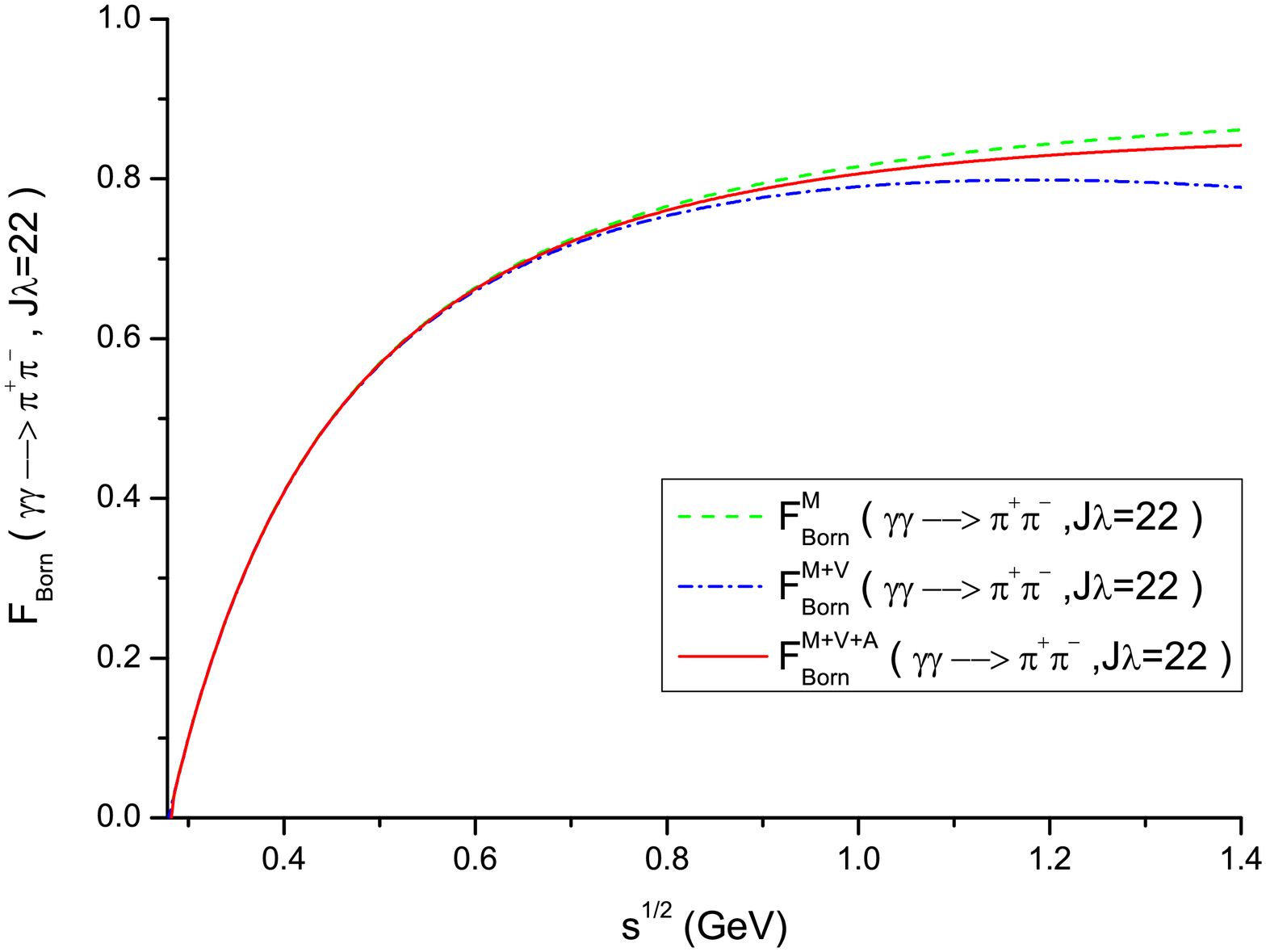}
\includegraphics[scale=0.23]{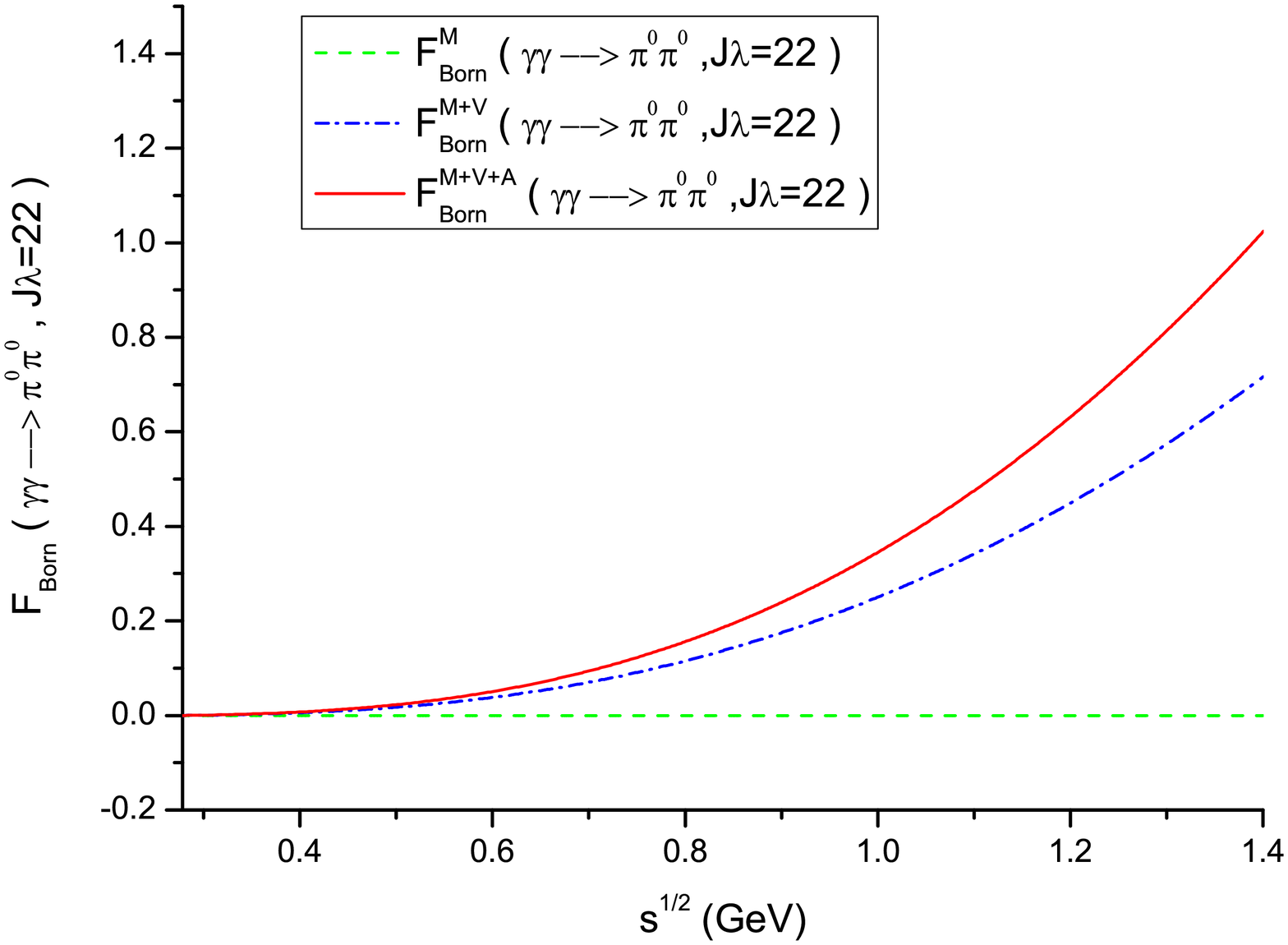}
\includegraphics[scale=0.23]{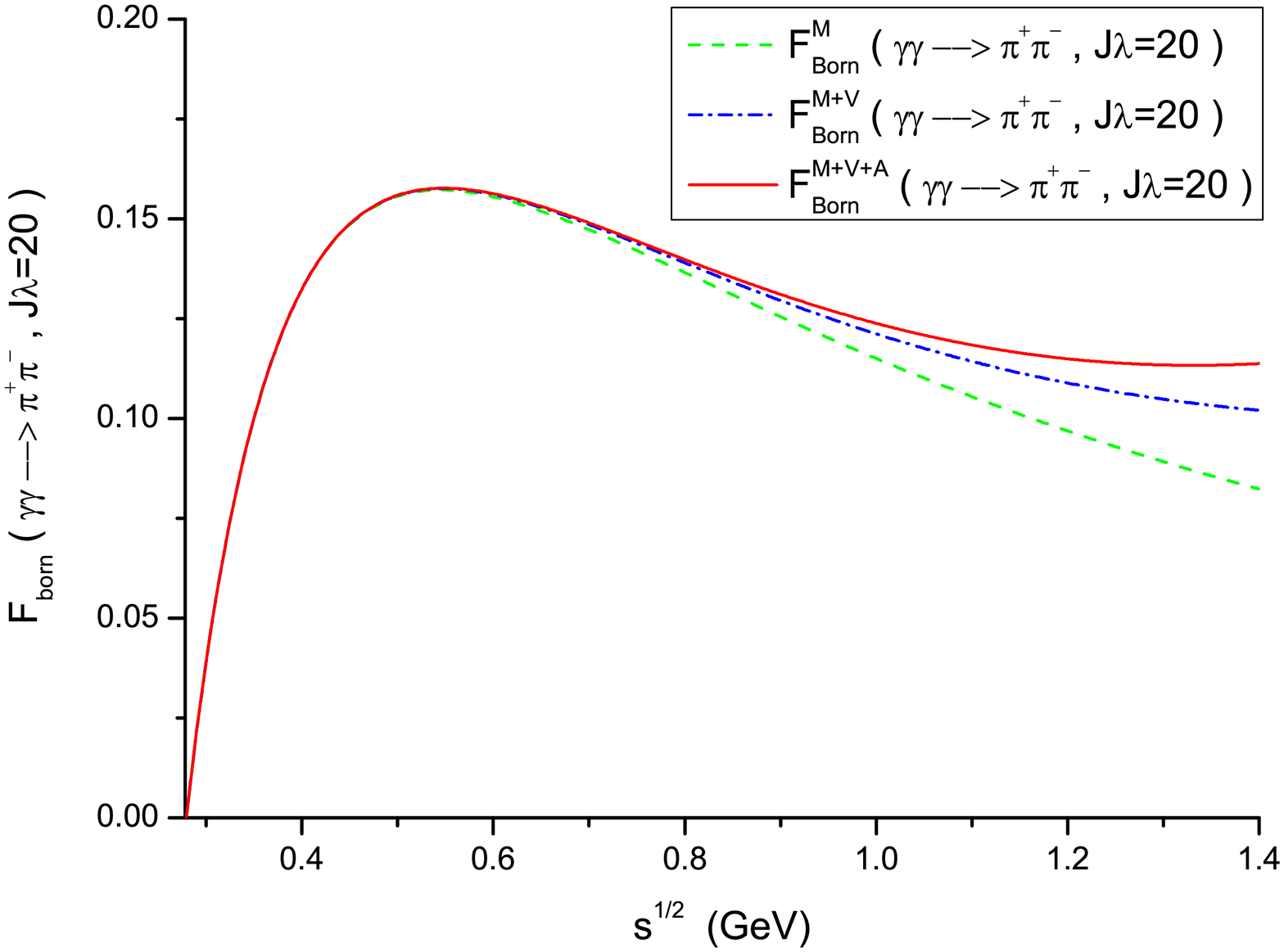}
\includegraphics[scale=0.23]{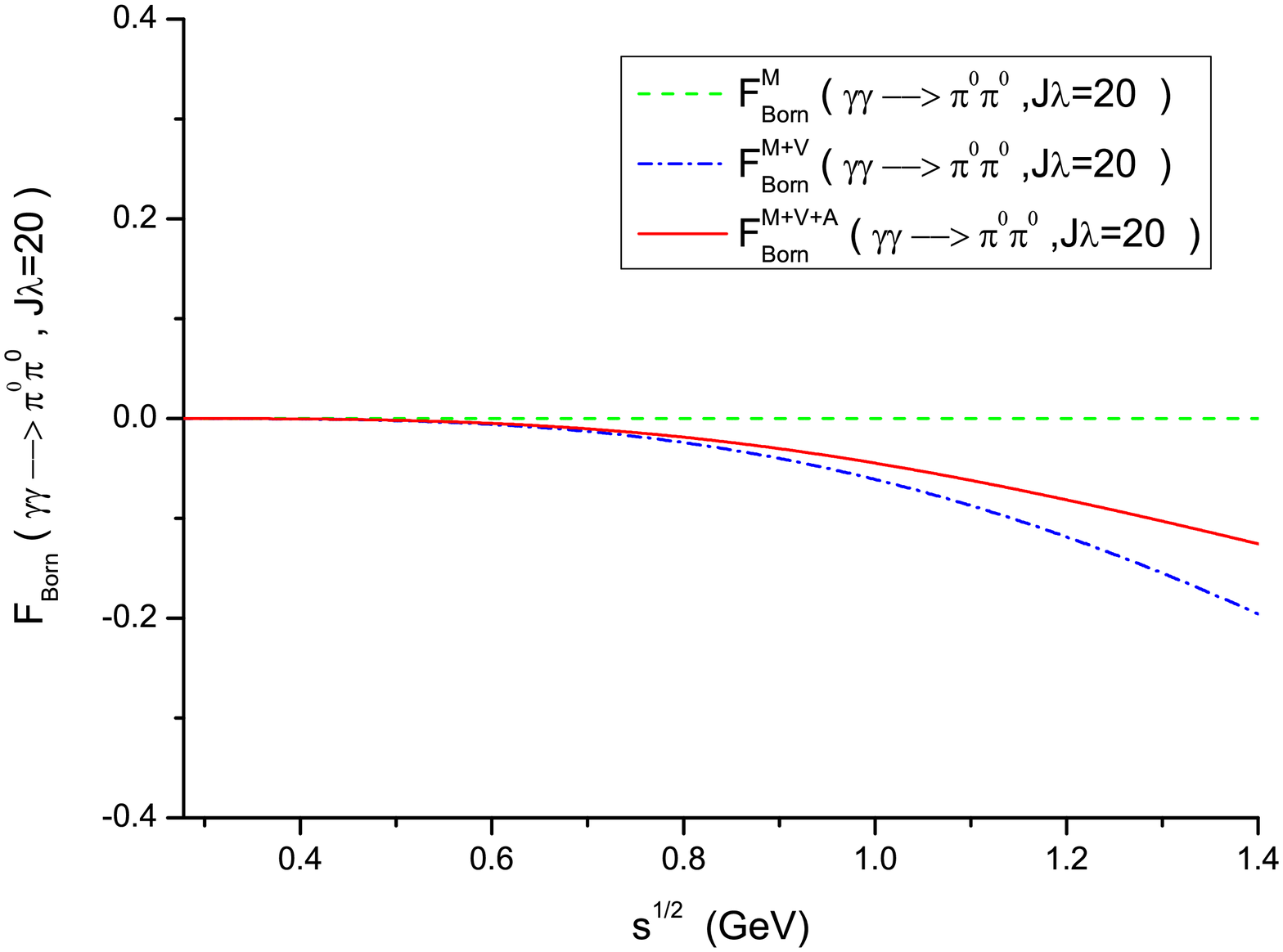}
\caption{\label{fig_Bornterm}Born term contribution to $\pi^+\pi^-$
and $\pi^0\pi^0$ amplitudes.  }
\end{figure}

Finally we plot in figure \ref{fig_Bornterm} various Born term
contributions in $\pi^+\pi^-$ and $\pi^0\pi^0$ channels.

\begin{thebibliography}{99}
\bibitem{belleExp}T.~Mori \textit{et al.} (Belle Collaboration), Phys. Rev.
{\bf D75}(2007)051101.

 \bibitem{Penn08}
 M.~R.~Pennington, T.~Mori, S.~Uehara, Y.~Watanabe,
Eur. Phys. J.{\bf C56}(2008)1.

 \bibitem{Penn072}
 M.~R.~Pennington, invited talk at YKIS Seminar on \textit{New Frontiers in QCD: Exotic Hadrons and Hadronic Matter},
 Kyoto, Japan, 20 Nov - 8 Dec 2006.
 Prog. Theor. Phys. Suppl. {\bf 168}(2007)143.


\bibitem{Morgan90}D.~Morgan, M.~R.~Pennington, Z.~Phys. {\bf
C48}(1990)623; D.~Morgan, M.~R.~Pennington,  Z.~Phys. {\bf
C37}(1988)431.

\bibitem{babelon76}O.~Babelon et al.,  Nucl. Phys.
{\bf B113}(1976)445; O.~Babelon et al.,  Nucl. Phys. {\bf
B114}(1976)252.


\bibitem{Mennessier83}G.~Mennessier, Z.~Phys. {\bf C16}(1983)241.

\bibitem{Anisovich99}A.~V.~Anisovich, V.~V.~Anisovich, Phys. Lett. {\bf B467}(1999)289.

\bibitem{Filkov05}L.~V.~Fil'kov, V.~L.~Kashevarov, Phys. Rev. {\bf C72}(2005)035211.

\bibitem{achasov07}N.~N.~Achasov, G.~N.~Shestakov, arXive:0712.0885
[hep-ph].



\bibitem{Oller07}J.~A.~Oller, L.~Roca, C.~Schat, Phys. Lett. {\bf B659}(2008)201.

\bibitem{prades}J.~Bernabeu, J.~Prades,  Phys. Rev. Lett. {\bf 100}(2008)241804.

\bibitem{Mennessier08}G.~Mennessier, S.~Narison, W.~Ochs, Phys. Lett. {\bf B665}(2008)205.


\bibitem{AMP}K.~L.~Au, D.~Morgan and M.~R.~Pennington, Phys. Rev.
{\bf D35}(1987)1633.

\bibitem{newke4} S.~Pislak \textit{et al}.,
Phys. Rev. {\bf D67}(2003)072004.

\bibitem{NA48}J.~R.~Batley et al (The NA48/2 Collaboration), Euro. Phys. J. {\bf C54}(2008)411.



\bibitem{Low54} F.~E.~Low
 Phys. Rev. {\bf 96}(1954)1428;
M.~Gell-Mann, M.~L.~Goldberger,
Phys. Rev. {\bf 96}(1954)1433;
 H.~D.~I.~Abarbanel, M.~L.~Goldberger,
 Phys. Rev. {\bf 165}(1968)1594.

\bibitem{Donoghue}J.~F.~Donoghue and Barry R. Holstein, Phys. Rev. {\bf
D48}(1993)137; D. Morgan and M. R. Pennington, Phys. Lett. {\bf
B272}(1991)134

\bibitem{donoghue2}J.~F. Donoghue, J.~Gasser, H.~Leutwyler, Nucl. Phys. {\bf B343}(1990)341.

\bibitem{pku3}Z.~Y.~Zhou \textit{et al}., JHEP 0502(2005)043.

\bibitem{pipiDwave}J.~J.~Wang, Z.~Y.~Zhou, H.~Q.~Zheng, JHEP {\bf 0512}(2005)019.

\bibitem{Ochs}W.~Ochs, Ph.D. thesis, Munich Univ., 1974.

\bibitem{cohen80}D.~H.~Cohen  \textit{et al}.,  Phys. Rev. {\bf D22}(1980)2595.

\bibitem{Etkin:1981sg}A.~Etkin  \textit{et al}.,  Phys. Rev. {\bf D25}(1982)1786 .

\bibitem{Martin:1979gm}A. D. Martin, E. N. Ozmutlu,  Nucl. Phys. B158(1979)520.

\bibitem{Costa:1980ji}  G.~Costa \textit{et al}.,  Nucl. Phys. {\bf
B175}(1980)402.

\bibitem{Polychronakos:1978ur} V.~A.~Polychronakos
 \textit{et al}.,  Phys.~Rev. {\bf D19}(1979)1317.
\bibitem{markushin}M.~P.~Locher, V.~E.~Markushin, H.~Q.~Zheng, Eur.~Phys.~J. {\bf C4}(1998)317.



\bibitem{Morgan92:pole-counting}
D.~Morgan, Nucl.~Phys.\ {\bf A543}(1992)632.

\bibitem{Behrend1992} H.~J.~Behrend \textit{et al}. (CELLO Collaboration), Z.~Phys. {\bf C56}(1992)381.

\bibitem{Boyer1990} J.~Boyer \textit{et al}. (Mark-II Collaboration), Phys.~Rev. {\bf D42}(1990)1350.

\bibitem{Marsiske1990} H.~Marsiske \textit{et al}.(Crystal Ball Collaboration) Phys. Rev. {\bf D41}(1990)3324.

\bibitem{Bienlein} J.~K.~Bienlein \textit{et al}, $IX$ $International$
$Workshop$ $on$ $Photon$-$Photon$ $Collisions$, San. Diego, 1992,
ed. D. Caldwell and H. P. Parr (World Scientific, 1992) p.241.

\bibitem{Li91} F.~E.~Close, Z.~P.~Li and T.~Barnes, Phys.~Rev. {\bf
D43}(1991)2161.



\bibitem{Pennington98} M.~Boglione, M.~R.~Pennington, Eur.~Phys.~J. {\bf
C9}(1999)11.

\bibitem{CCL}I.~Caprini,
G.~Colangelo, H.~Leutwyler,   Phys. Rev. Lett. 96, 132001 (2006).

\bibitem{kaminski}R.~Kaminski G.~Mennesier, S.~Narison, to appear.
\bibitem{kato}M.~Kato, Ann. Phys.{\bf 31}(1965)130.
\bibitem{Fujii}Y.~Fujii, M.~Kato,Nuovo Cimento {\bf 13A}(1973)311.

\bibitem{Bugg93} B.~S.~Zou, D.~Bugg, Phys. Rev. {\bf D48}(1993)3948.


\bibitem{pku1}
 Z.~Y.~Zhou, H.~Q.~Zheng, Nucl.
Phys.{\bf A775}(2006)212;  H.~Q.~Zheng et al., Nucl. Phys. {\bf
A733}(2004)235;




 \bibitem{Rosner76} J.~Babcock, J.~L.~Rosner, Phys. Rev. {\bf D14}(1976)1286.

 \bibitem{Barnes85} T.~Barnes, Phys. Lett. {\bf B165}(1985)434.
\bibitem{Narison06}S.~Narison, Phys. Rev. {\bf D73}(2006)114024; S. Narison, G. Veneziano,
Int. J. Mod. Phys. {\bf A4}(1989)2751; S.~Narison, Nucl. Phys. {\bf
B509}(1998)312.
 \bibitem{Achasov82}
N.~N.~Achasov \textit{et al}.,  Z.~Phys. {\bf C16}(1982)55.

 \bibitem{Barnes92}
T.~Barnes, Proc. \textit{IXth Int. Workshop on Photon-Photon
Collisions} (San Diego, 1992), ed. D.~Caldwell and H.~P.~Paar(World
Scientific, 1992.), p.263

 \bibitem{Hanhart07} C.~Hanhart, Yu.~S.~Kalashnikova,
A.~E.~Kudryavtsev, A.~V.~Nefediev, Phys. Rev. {\bf D75}(2007)074015.


\bibitem{PDG2008}  C.~Amsler et al. (Particle Data Group), Phys. Lett. {\bf B667}(2008)1.

\bibitem{Pennington07}M.~R.~Pennington,  Mod. Phys. Lett. {\bf
A22}(2007)1439.

\bibitem{Giacosa}F.~Giacosa, T.~Gutsche, V.~Lyubovitskij, Phys. Rev. {\bf D77}(2008)034007.

\bibitem{zheng08}
H.~Q.~Zheng, Plenary talk given at \textit{Workshop on Chiral
Symmetry in Hadron and Nuclear Physics: Chiral07}, Osaka, Japan,
13-16 Nov 2007; Mod. Phys. Lett. {\bf A23}(2008)2218.

\bibitem{guo06}
Z.~H.~Guo, L.~Y.~Xiao, H.~Q.~Zheng, Int. J. Mod. Phys. {\bf
A22}(2007)4603; M.~X.~Su, L.~Y. Xiao, H.~Q.~Zheng, Nucl. Phys. {\bf
A792}(2007)288.





\bibitem{Gasser94} J.~Gasser, S.~Bellucci, M.~E.~Sainio, Nucl. Phys. {\bf
B423}(1994)80.

\bibitem{Suzuki} M.~Suzuki,  Phys. Rev. {\bf D47}(1993)1252.

\bibitem{Cheng} H.~Y.~Cheng, Phys. Rev. {\bf D67}(2003)094007.

\bibitem{Guo}Z.~H.~Guo, Phys. Rev. {\bf D 78}(2008)033004.

\end{thebibliography}
\end{document}